\documentclass[twocolumn,times,astrosymb,floatfix]{aastex631}

\usepackage{
	amsmath,
	amssymb,
	amsthm,
	booktabs,
        cancel,
	epsfig,
	graphicx,
	grffile,
	import,
	morefloats,
	microtype,
	natbib,
	subfigure,
	ulem,
	url,
	verbatim,
	wasysym,
	xcolor,
	xspace,
}
\usepackage{soul}

\usepackage{savesym}
\savesymbol{tablenum}
\usepackage{siunitx}
\restoresymbol{SIX}{tablenum}
\usepackage[version=4]{mhchem}

\DeclareSIUnit{\jy}{Jy}
\DeclareSIUnit{\beam}{beam}
\DeclareSIUnit{\kms}{\kilo\meter\per\second}

\shorttitle{Vertical Structure of Gas and Dust in Four Debris Disks}
\shortauthors{Worthen et al.}

\begin{document}

\title{\Large
Vertical Structure of Gas and Dust in Four Debris Disks
}

\correspondingauthor{Kadin Worthen}
\email{kworthe1@jhu.edu}

\author[0000-0002-5885-5779]{Kadin Worthen}
\affiliation{William H. Miller III Department of Physics and Astronomy, John's Hopkins University, 3400 N. Charles Street, Baltimore, MD 21218, USA}

\author[0000-0002-8382-0447]{Christine H. Chen}
\affiliation{Space Telescope Science Institute, 3700 San Martin Drive, Baltimore, MD 21218, USA }
\affiliation{William H. Miller III Department of Physics and Astronomy, John's Hopkins University, 3400 N. Charles Street, Baltimore, MD 21218, USA}

\author[0000-0001-5638-1330]{Sean Brittain}
\affiliation{Department of Physics and Astronomy, 118 Kinard Laboratory, Clemson University, Clemson, SC 29634, USA}

\author[0000-0001-9352-0248]{Cicero Lu}
\affiliation{William H. Miller III Department of Physics and Astronomy, John's Hopkins University, 3400 N. Charles Street, Baltimore, MD 21218, USA}

\author[0000-0002-4388-6417]{Isabel Rebollido}
\affiliation{Centro de Astrobiología (CAB), INTA-CSIC, Camino Bajo del Castillo s/n - Villafranca del Castillo, 28692 Villanueva de la Cañada, Madrid, Spain }

 \author{Aoife Brennan}
 \affiliation{School of Physics, Trinity College Dublin, the University of Dublin, College Green, Dublin 2, Ireland}

 \author[0000-0003-4705-3188]{Luca Matr\`a}
 \affiliation{School of Physics, Trinity College Dublin, the University of Dublin, College Green, Dublin 2, Ireland}

 \author[0000-0001-9834-7579]{Carl Melis}
 \affiliation{Center for Astrophysics and Space Sciences, University of California, San Diego, CA 92093-0424, USA}
 
  \author{Timoteo Delgado}
 \affiliation{Center for Astrophysics and Space Sciences, University of California, San Diego, CA 92093-0424, USA}
 
 \author[0000-0002-2989-3725]{Aki Roberge}
 \affiliation{Exoplanets and Stellar Astrophysics Lab, NASA Goddard Space Flight Center, Greenbelt, MD 20771, USA}

 \author[0000-0002-9133-3091]{Johan Mazoyer}
 \affiliation{7 LESIA, Observatoire de Paris, Université PSL, CNRS, Université Paris Cité, Sorbonne Université, 5 place Jules Janssen, F-92195 Meudon, France}

\begin{abstract}
 We present high-spectral resolution M-band spectra from iSHELL on NASA's Infrared Telescope Facility (IRTF) along the line of sight to the debris disk host star HD 32297. We also present a Gemini Planet Imager (GPI) H-band polarimetric image of the HD 131488 debris disk. We search for fundamental CO absorption lines in the iSHELL spectra of HD 32297 but do not detect any. We place an upper limit on the \ce{CO} column density of $\sim$6$\times10^{15}$ cm$^{-2}$. By combining the column density upper limit, the \ce{CO} mass measured with ALMA, and the geometrical properties of the disk, we estimate the scale height of the \ce{CO} to be $\lesssim$ 2 au across the radial extent of the disk ($\sim$80-120 au). We use the same method to estimate the \ce{CO} scale height of three other edge-on, CO-rich debris disks that all have \ce{CO} observed in absorption with HST as well as in emission with ALMA: $\beta$ Pictoris, HD 110058, and HD 131488. We compare our estimated CO scale heights of these four systems to the millimeter dust scale heights and find that, under the assumption of hydrostatic equilibrium, there is a potential correlation between the CO and millimeter dust scale heights. There are multiple factors that affect the gas vertical structure such as turbulence, photodissociation with weak vertical mixing, as well as where the gas originates. One possible explanation for the potential correlation could be that the gas and dust are of a similar secondary origin in these four systems. 
\end{abstract}

\keywords{
    planet formation ---
    debris disks ---
    circumstellar matter ---
    infrared astronomy 
}


\section{Introduction}\label{sec:Introduction}
Debris disks are dust-dominated planetary systems consisting of planets, planetesimals, and dust. The dust is thought to be continuously replenished through collisions between planetesimals that have been perturbed into eccentric and inclined orbits by planets within the disk \citep{wyatt2008,hughes18}. Classically, debris disks are expected to be gas free, but millimeter observations in the past decade have found \ce{CO} to be present in $\sim$20 debris disk systems (e.g. \citealt{sifry16,marino16,131488_alma,32297_ALMA,rebollido22}.) If the planetesimals in debris disks contain ice, then their destruction could liberate gas phase molecules into the circumstellar environment and the gas and the dust would originate from the same population of planetesimals \citep{kral17}. However, the gas present in debris disks could also be remnant of the protoplanetary disks, as long gas densities are high enough that shielding from UV photons significantly increases the photodissociation lifetime of \ce{CO} (e.g. \citealt{Kospal13}, \citealt{Nak21}). Whether the gas in debris disks is remnant from the protoplanetary disk or of second generation, released from collisions between minor bodies, is still an open question, especially for the most gas rich debris disks. 

Sub-mm observations of debris disks from observatories like the Atacama Large
Millimeter/submillimeter Array (ALMA) have found $\sim$ 20 systems that posses CO gas (e.g. \citealt{Kospal13,131488_alma,bpic_almaco,32297_ALMA}). These observations typically trace cold gas that is several tens of au away from the central star \citep{marino16,sifry16,bpic_almaco}. Across the population of debris disks with cold gas detected, the measured CO masses span a range of 5 orders of magnitude from $\sim10^{-6} M_{\oplus}$ to $\sim10^{-1} M_{\oplus}$ \citep{kral17,moor19}. The debris disks that are the most CO-rich are found around young (10-50 Myr) A-type stars \citep{Kospal13,moor19}, while lower CO mass debris disks ($<10^{-4} M_{\oplus}$) have been found around a range of spectral types from G to A-type stars \citep{sifry16}. 

A limiting factor to the amount of \ce{CO} in a debris disk is the photodissociation timescale due to interstellar UV photons. For unshielded \ce{CO} molecules their photodissociation lifetime is $\sim$120 years \citep{visser09}. This short lifetime makes it difficult to explain the origin of the gas in debris disks with large \ce{CO} masses using the model of second generation gas production \citep{disk_gassample}. Recent theoretical studies have shown that neutral atomic carbon gas (CI) produced through the photodissociation of \ce{CO}, can provide shielding from  UV photons, prolonging the CO gas lifetime in debris disks \citep{kral2019,marino20}. This neutral carbon does not experience significant radiation pressure from the central star and can accumulate over time \citep{Fernandez2006, Brandekar11}. Similarly, neutral oxygen, another photodissociation product of CO, does not experience significant radiation pressure from the star. Thus, carbon and oxygen are expected to be the dominant gas species in CO rich debris disks (e.g.\citealt{Roberge2006}, \citealt{Brandekar11}). At high gas densities, self shielding can also extend the photodissociation lifetime of CO and allow the number of CO molecules to build over time \citep{kral2019}. These shielding scenarios can allow for second generation \ce{CO} to accumulate to larger masses ($\sim10^{-2}-10^{-1}M_{\oplus}$). While these studies show that high \ce{CO} mass of secondary origin is possible in debris disks, they are not able to rule out the possibility of the gas being of primordial origin. 

 CO-rich debris disks that are edge-on (\textit{i}$>80$ degrees) provide a unique opportunity to study cold CO gas in both emission at sub-mm wavelengths and absorption at infrared fundamental band (M-band) and UV A-X band. Four such CO-rich, edge-on debris disks exist:  HD 32297, $\beta$ Pictoris, HD 110058, and HD 131488 \citep{32297_ALMA,131488_alma,bpic_almaco}. Three of these systems ($\beta$ Pic, HD 110058, and HD 131488) have \ce{CO} detected using absorption line spectroscopy (\citealt{Roberge_2000}, \citealt{bpicabsorp}, Brennan submitted). All four systems have detections of volatile atomic gas (e.g. C and O) either in absorption with HST (\citealt{Roberge2006}, Brennan et al. submitted) or in emission with Herschel or ALMA (\cite{Brandeker16}, \cite{cataldi20}). There is potentially another similar system, 49 Ceti (A1V, $\sim$ 40Myr; \citealt{49cet88,zuckerman12}) that is not discussed in our analysis. 49 Ceti is a CO-rich debris disk seen in emission with ALMA \citep{nhung17}, but has a non-detection of CO in absorption with HST \citep{roberge14}. 49 Ceti, however, is not as edge-on as the other four, as its disk has a measured inclination of 79 degrees in thermal emission with ALMA \citep{hughes17}. 
 
By combining column density measurements from absorption line spectroscopy and CO mass measurements from emission at millimeter wavelengths, we can constrain the vertical distribution of the gas provided that the geometry of the disk is known (inclination, inner and outer radii). The 4 systems listed above are thus ideal for this measurement of gas vertical structure. Currently, $\beta$ Pic is the only one of these four systems that has a measurement of the CO scale height from ALMA \citep{bpic_almaco}. 

The vertical structure of a debris disk is an important observable because it may reflect the inclination distribution of planetesimals in a system. The vertical distribution of the planetesimals has been suggested to be a probe of the velocity dispersion of the solid bodies in a disk and therefore the dynamical excitation from potentially unseen planetary  mass companions \citep{Brown2001, hughes18}. Thus, the vertical structure of extra solar planetesimal belts can provide evidence for the gravitational effects of unseen planets. Millimeter-sized dust probed by ALMA is expected to trace the location of the parent planetesimal belt, as particles of these sizes are not strongly affected by radiation forces \citep{wyatt2008}. The micron-sized dust grains in debris disks traced by scattered light images from high-contrast imaging instruments like GPI and SPHERE are affected more strongly by radiation forces and will migrate outwards beyond the parent planetesimal belt with time \citep{wyatt2008}. Radiation forces can also increase the orbital inclinations of micron-sized dust particles \citep{Thebault09}, so the vertical distribution of the micron-size dust does not trace the vertical distribution of the parent planetesimal belt as well as the millimeter-sized dust. It is thus expected, if the \ce{CO} in debris disks originates from collisions within the same planetesimal belt as the dust, that the \ce{CO} and millimeter-sized dust should have similar spatial distributions and vertical structures \citep{Kospal13,hughes18}. Measuring the spatial distribution of the millimeter-sized dust and gas is a potential method to investigate the origin of gas in debris disks.

In this paper, we present IRTF iSHELL observations looking for fundamental ro-vibrational transitions of \ce{CO} in absorption towards HD 32297 in M-band that resulted in a non-detection. We also present GPI H-band polarimetry observations of the HD 131488 debris disk. We constrain the scale height of the gas in HD 32297 and three other debris disks in section 3. In section 4 we discuss our findings about the vertical structure of the gas and dust in these four systems and in section 5 we conclude and summarize.

\section{Observations and Analysis}\label{sec:Observations}
\subsection{iSHELL Observations}
\begin{figure*}[!htbp]
    \centering
    \includegraphics[scale=0.7]{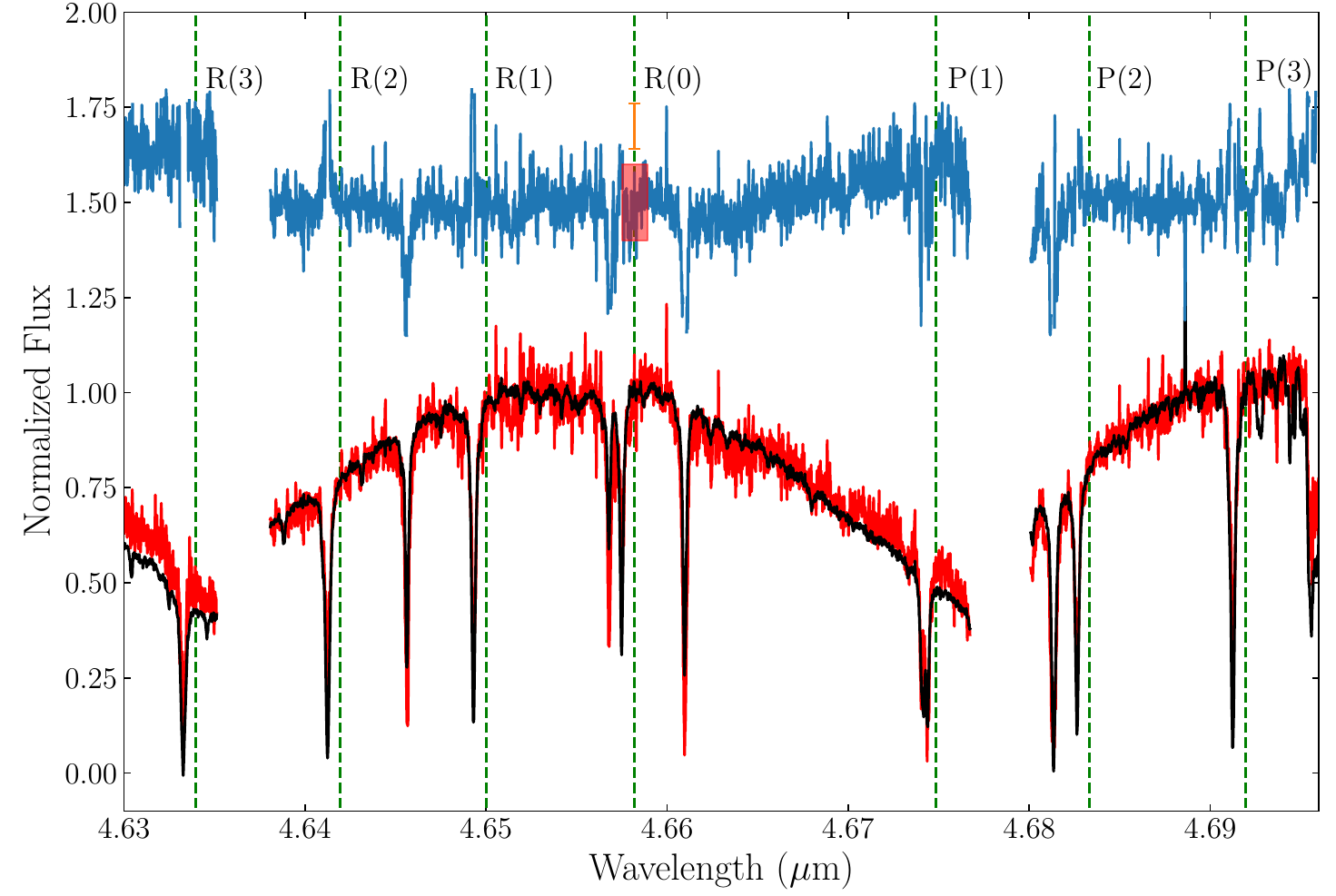}
    \singlespace\caption{Telluric corrected iSHELL spectrum of HD 32297 from 4.63-4.70 $\text{\textmu}$m (blue) in normalized flux units. The telluric standard (black) and the not telluric corrected HD 32297 spectrum (red) are also plotted in normalized units. The telluric corrected HD 32297 spectra is offset by 0.5 units. The dashed green vertical lines show the expected locations of the CO v=1-0 R(1)-R(3) and P(1)-P(3) CO lines after correcting for the Doppler shift created by the radial velocity of the star and Earth's orbital motion. We do not detect any of the CO lines in absorption. The orange error bar shows the standard deviation of the 50 data points around the R(0) CO line and is offset from the spectrum vertically for clarity. The red box shows the region used to calculate the standard deviation.}
    \label{fig:ishellspec}
\end{figure*}

We used the iSHELL instrument, an immersion grating echelle
spectrograph, \citep{iShells} on the NASA IRTF to search for fundamental CO absorption towards the HD 32297 debris disk. We observe our science target star HD 32297 (A0V, K=7.594) first followed immediately after by observations of the nearby (6 degrees) bright star HD 35468 (B2V, K=2.32), which we used to remove the telluric lines from the HD 32297 spectrum. A total of 107 minutes of data were taken for HD 32297 using the M1 grating setting (4.52-5.25 $\text{\textmu}$m) with a slit-width of 0.75", providing a spectral resolution of R$\sim$60,000 \citep{banzatti22}. This wavelength range and spectral resolution allows us to search for fundamental R-branch and P-branch ro-vibrational absorption lines from circumstellar CO. The average airmass across the total integration time on HD 32297 was 1.1. We used the same observational setup as HD 32297 to observe HD 35468 and the total integration time for HD 35468 was 56 seconds. The average airmass across the total integration time for HD 35468 was also 1.1. All the observations were taken with an ABBA nod pattern to remove the thermal sky background to first order.

We reduce the data using Spextool v5.0.3 \citep{spextool,telluric}. The reduction process includes flat fielding, non-linearity correction, pair subtraction, aperture extraction, telluric correction, and wavelength calibration. The wavelength calibration is done using telluric features and an atmosphere transmission model described in \cite{spextool}. For spectral extraction, we use an aperture radius equal to 1.0 arcseconds. To correct the telluric lines from the target spectrum, we divide the spectrum of HD 32297 by that of our telluric calibrator star. The spectrum of HD 32297 before the telluric correction has a signal-to-noise ratio of $\sim$21 on order 111 (4.63-4.68 $\text{\textmu}$m), which contains the CO ro-vibrational lines of interest. The final telluric corrected spectrum of HD 32297 has a signal-to-noise ratio of $\sim$20 across 4.63-4.68 $\text{\textmu}$m and the spectrum of the telluric standard HD 35468 has a signal-to-noise ratio of $\sim$60 over the same wavelength range. 

We do not detect the CO R(0), R(1), R(2), R(3) or the P(1), P(2), P(3) lines in absorption toward HD 32297. The portion of the iSHELL spectrum of HD 32297 that contains these CO transitions, from 4.63-4.70 $\text{\textmu}$m, is shown in Figure \ref{fig:ishellspec}. The green dashed lines show the expected positions of each of the fundamental CO lines after correcting for the Doppler shift due to Earth's orbital velocity on the night of the observations and the radial velocity of HD 32297, which is 20.6 km s$^{-1}$ in the barycentric frame \citep{32297_ALMA}. HD 32297 has an LSRK velocity of 5.30$\pm$0.1 km s$^{-1}$.

We use the non-detection of the CO R(0) absorption line in the R$\sim$60,000 iSHELL spectrum to place an upper limit on the circumstellar CO column density of the HD 32297 debris disk. We assume a Gaussian line profile with a depth equal to the standard deviation (1$\sigma$) of the 50 data points ($\Delta\lambda=0.0007$ $\text{\textmu}$m) centered on the expected location of the R(0) CO line ($\sim$4.66 $\text{\textmu}$m), as shown by the orange point in Figure \ref{fig:ishellspec}. This 1$\sigma$ limit of the R(0) line is 0.055 in units of normalized flux and is used to compute the 3$\sigma$ upper limit on the equivalent width and column density. We use the average FWHM (0.0003 $ \mu$m) of the unsaturated telluric lines as the width of the CO lines in calculating the upper limit on equivalent width since we expect the CO lines to be spectrally unresolved. 

 We estimate a 3$\sigma$ upper limit on the equivalent width of the  R(0) line to be 1.4$\times10^{-5}$ $\text{\textmu}$m. We then calculate an upper limit on CO column density in the $i$th state using the relation:
\begin{equation}
N_{i}=\frac{W}{8.85\times10^{-13}f_{ij}\lambda{^2}}
\end{equation}
where $W$ is the equivalent width of the line, $8.85\times10^{-13}$ cm is the classical electron radius, $\lambda$ is the wavelength of the line, and $f_{ij}$ is the oscillator strength of the transition. See table \ref{tab:line_params} for the line parameters used to calculate the column density. Assuming that the CO is in LTE, the total column density of CO is given by 
\begin{equation}
    N_{tot}=\frac{N_{i}Qe^{E_i/kT}}{g_i}
\end{equation}
where $N_i$ is the column density of the ith state, $Q$ is the ro-vibrational partition function, $E_i$ is the energy of the $i$th state, $T$ is the gas temperature, and $g_i$ is the statistical weight of the $i$th state. We use a temperature of 30 K for the column density upper limit calculation, which is the temperature that was used for the CO mass calculation from the ALMA observations of HD 32297 system \citep{32297alma}. At 30 K, the 3-$\sigma$ CO column density upper limit using the R(0) ($\sim$4.66 $\text{\textmu}$m) CO absorption line towards HD 32297 is $6\times10^{15}$ cm$^{-2}$. For comparison, the line of sight to $\beta$ Pic has a measured CO column density of 2$\times10^{15}$ cm$^{-2}$ in the infrared \citep{bpicabsorp} and 6$\times10^{14}$ cm$^{-2}$ in the UV \citep{Roberge_2000}.

We also estimate the column density upper limit of CO without assuming that the gas is in LTE, since this assumption may be incorrect. Instead, we follow an approach similar to that in \cite{bpicabsorp}, where the total column density of CO seen in absorption in the $\beta$ Pic disk was equal to the sum of the measured column density of the R(0), R(1), and R(2) lines. This was because the CO lines were optically thin at mid-infrared wavelengths, the R(2) line was sub-thermally populated, and no other higher energy transitions of CO were detected, indicating that the gas was too cold to populate higher energy states. So the total column density of CO towards $\beta$ Pic in the mid-IR was equal to the sum of the column density of those three transitions. For HD 32297, we also expect the gas to be cold and only the lowest energy levels to be populated like in $\beta$ Pic because both $\beta$ Pic and HD 32297 are similar spectral type and both disks have similar radial distributions (see Table \ref{tab:disk_prop}). We estimated the equivalent width upper limits of the R(1) and R(2) lines with the same method as the R(0) line described above. The line properties used in this calculation are shown in Table \ref{tab:line_params}. The 3$\sigma$ equivalent width upper limits for the R(1) and R(2) lines are 3.9$\times10^{-5}$ $\text{\textmu}$m and 3.1$\times10^{-5}$ $\text{\textmu}$m respectively. We then used Equation 1 to calculate the 3$\sigma$ column density upper limit for all three CO lines and summed all three upper limits together to get a total CO column density upper limit. This gives a total CO column density upper limit of 7$\times10^{15}$ cm$^{-2}$ for HD 32297. This is similar to the column density upper limit inferred from assuming the gas is in LTE (6$\times10^{15}$ cm$^{-2}$), so the method for calculating the column density upper limit for HD 32297 does not affect the following analysis. 

\begin{table}[!htpb]
    \centering
    \caption{CO Line Parameters}
    
    \begin{tabular}{c c c c}
    \hline
          Line & Wavelength ($\mu m$)& $f$ 
          (10$^{-6}$) &Column Density (cm$^{-2}$)  \\
    \hline
          R(0)& 4.6575 &11.420 & $<2\times10^{15}$ \\
          R(1)& 4.6493 &7.6267 & $<3\times10^{15}$ \\
          R(2)& 4.6412 &6.8760 & $<2\times10^{15}$ \\
    \hline
    
    \end{tabular}
    \begin{minipage}{8cm}
    \vspace{0.1cm}
        Note: Oscillator strengths ($f$) for each transition are taken from \cite{CO_rovib}.
    \end{minipage}
    
    \label{tab:line_params}
\end{table}

\subsection{GPI observations}
 We obtained GPI \citep{Macintosh14} observations of the HD 131488 debris disk in H-band using polarimetry mode on September 1, 2015. Our observing sequence included a total of fifty-eight 60 second exposures with a seeing of 0.65 arcseconds and a mean airmass of 1.21. Throughout the imaging sequence, the half-wave plate (HWP) was rotated between position angles of 0, 22.5, 45, and 67.5 degrees. 

\begin{figure}[!htbp]
    \centering
    \includegraphics[scale=0.35]{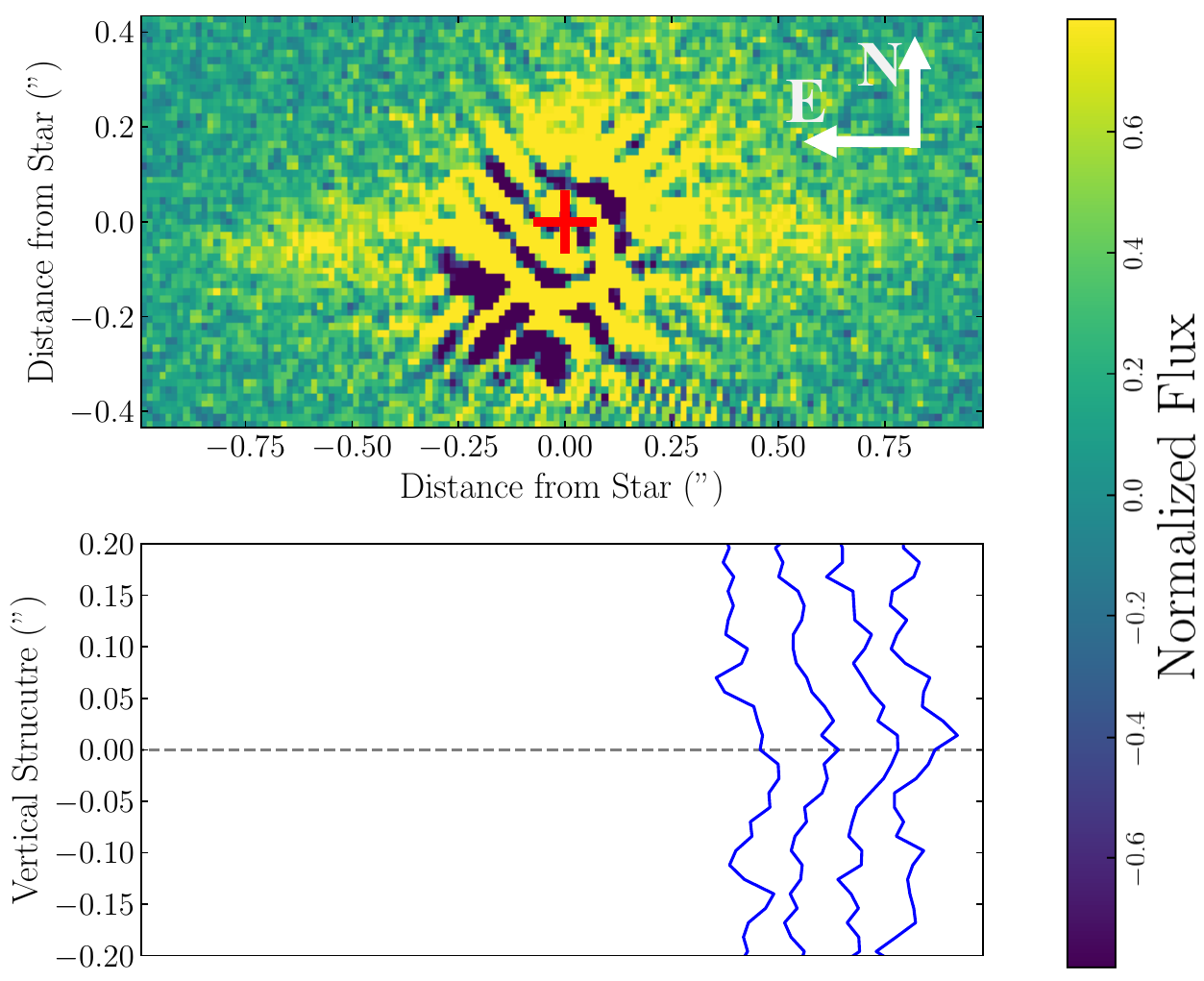}
    \singlespace\caption{ Top: Rotated H-band Stokes $Q_r$ image of HD 131488 showing an edge-on debris disk. The location of the star is shown by the red cross. Bottom: Vertical cuts of every 10 pixel columns averaged along the radial direction of the disk midplane of the GPI image from the west side (see Section 2.2). These vertical cuts were averaged to compare with the GPI PSF in H-band (see Figure 3). The black dashed line shows the vertical location of the star in the image.}
    \label{fig:GPI_image}
\end{figure}

We reduced the observations with the GPI Data Reduction Pipeline (DRP) \citep{Perrin14} following the same procedure as outlined in \cite{Perrin15}. The data reduction involved dark subtraction, correction for instrument flexure, microphonics noise, and bad pixels. Each frame was then assembled into a polarization data cube, including two image slices, one of which was the left-handed circularly polarized light and the other the right-handed circularly polarized light. We then used Polarized Differential Imaging (PDI), where the two orthogonal polarization states are differenced at the various HWP rotation angles to cancel out the unpolarized light from the central star. This results in a Stokes data cube with the \textit{I,Q$_r$,U$_{\phi}$} polarization states. The final GPI Stokes $Q_r$ image of the HD 131488 debris disk produced from PDI is shown in Figure \ref{fig:GPI_image}. 

 We measured the vertical structure of the HD 131488 debris disk in scattered light from the GPI H-band polarimetry mode image and compared it to the GPI PSF at H-band. To represent the GPI PSF, we used an archival unocculted H-band Integral Field Spectrograph (IFS) observation of the binary star system HD 6307 \citep{Derosa20} taken on the same night as the HD 131488 observations. We use the brightest of the two stars in the system (F5, H=7.65) to represent the PSF. We reduce the HD 6307 data using the GPI DRP following the procedure described in \cite{Perrin15} to produce a final H-band data cube. We estimated the profile of the GPI PSF at H-band by collapsing the IFS data cube along the wavelength axis and then taking a line cut through the PSF center. 
 
 To measure the scale height of the disk in the GPI image, we first rotated the image by 96 degrees, corresponding to the position angle measured by \cite{131488_alma}, to ensure that the midplane is aligned with the horizontal axis. We then took vertical cuts of the disk at each column of pixels from 0.25 to 0.7 arcseconds from the central star on the west side of the disk shown in Figure \ref{fig:GPI_image}. We only took vertical cuts of the west side of the disk because there appears to be a bend in the disk at around -0.5 arcseconds from the central star on the east side. We do not comment further on this apparent bend or whether it is statistically significant because that is beyond the scope of this paper. We only mention it here because we avoid this area when taking the vertical cuts, since it could affect the measurement of the vertical profile. Because the individual vertical cuts are low S/N, we averaged all the vertical cuts from the west side of the disk to get a higher S/N vertical profile. 
 
\begin{figure}[!htbp]
    \hspace{-2mm}
    \includegraphics[scale=0.55]{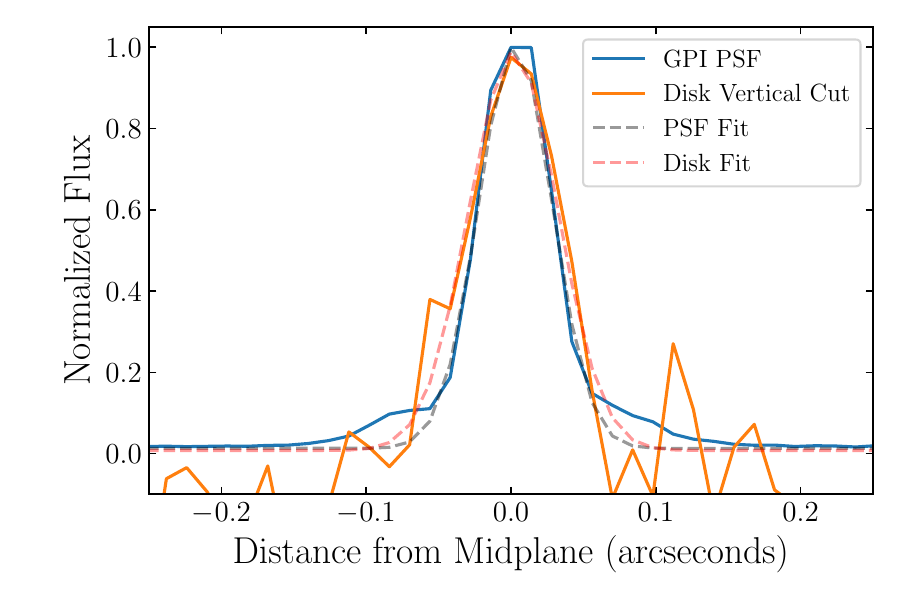}
    \singlespace\caption{Average of vertical cuts of HD 131488 GPI image (orange) with and a cut of the GPI PSF at H-band showing its resolution element (blue). The best fit Gaussian profiles of the PSF (black) and disk (red) are shown as dashed lines. The profiles are plotted such that the baseline of their Gaussian fits match. The vertical cut of the disk has a FWHM consistent with that of the GPI PSF, so the disk is vertically unresolved in scattered light in H-band.}
    \label{fig:131488_vert}
\end{figure}

As discussed in \citep{bpic_almaco}, vertical cuts observed on sky at any midplane location of the disk will tend to trace the vertical profile at the outer radius of the disk. By averaging the vertical profiles shown in Figure \ref{fig:GPI_image}, we are measuring the on-sky vertical profile at the outer radius of the disk and comparing this to the GPI PSF to check if it is vertically resolved. Averaging the vertical cuts is likely not biased by the vertical cuts at different disk radii, because all the vertical cuts will tend to trace the vertical structure at the outer disk radius. The vertical cuts are not shifted to be aligned with the disk spine before averaging because we are not able to fit for the spine of the disk with the low S/N individual vertical cuts. The average of the vertical cuts is shown in Figure \ref{fig:131488_vert}. 
 
 We measured the FWHM of the average of vertical cuts of the disk and the to be 0.07$\pm$0.01 arcseconds and the FWHM of the GPI PSF at H-band to be 0.060$\pm$0.006 arcseconds. The FWHM of the PSF of GPI at H-band is within the uncertainties of the FWHM of average of the vertical cuts of the disk. Therefore, we do not vertically resolve the HD 131488 debris disk with GPI, and we are only able to place an upper limit on its scale height in scattered light. At a distance of 152 parsecs \citep{GAIA} and with the angular resolution of GPI, the scale height upper limit for the HD 131488 debris disk in scattered light is $\sim$5-6 au. Assuming a disk outer radius of 115 au (see Section \ref{sec:alma_cont}), this gives an upper limit on the aspect ratio of $H/r\lesssim0.05$. 
 
 Before performing forward modeling of scattered light images, both \cite{Gaspar20} and \cite{Crotts21} found that, through a similar vertical cut analysis, the vertical profile of their respective disks was larger than the instrument PSF. Because we do not find that the disk vertical profile for HD 131488 is spatially resolved, we do not forward model the disk to measure the vertical profile and instead we report a scale height upper limit based on the the measured FWHM of the GPI PSF. Since the disk is unresolved vertically, forward modeling would not provide a better constraint on the scale height and would similarly just give an upper limit.

 \begin{table*}[!htbp]
     \centering
     \caption{Stellar parameters of the four targets}
     \begin{tabular}{c c c c c c c c }
     \hline
     Source  & $M_* (M_{\odot})$& Distance (pc)* & Age (Myr) & ALMA \textit{i} (\textdegree)& Scattered light \textit{i} (\textdegree) &$M_{CO} (M_{\oplus})$& references  \\
     \hline
          HD 32297 & 1.7& 133 & $<30$ & 88.4$\pm$0.1& 88.75$^{+0.02}_{-0.02}$&7.4$\pm1.3\times10^{-2}$& (1),(2),(14),(15),(17)  \\
          $\beta$ Pic & 1.8 & 19.4 &23-29 & 86.6$\pm0.4$& 85.27$^{+0.26}_{-0.19}$&3.4$\pm0.5\times 10^{-5}$& (3),(4),(5),(6), (13) \\
          HD 131488  & 1.9 &152 &10-20 & 85$^{+3}_{-2}$& NA&8.9$\pm1.5\times 10^{-2}$&(7),(8),(10),(16) \\
          HD 110058 & 1.8 &130 & 12-18 & 85.5$^{+2.5}_{-7.2}$ &$>80$& 0.07$^{+3.5}_{-0.06}$& (9),(11),(12),(16),(18)\\
        
     \hline
     \end{tabular}
      \begin{minipage}{17cm}
      \vspace{0.1cm}
      \textbf{Notes:} *Distances from: \cite{Gaia18} 
  References: (1) \cite{Kalas2005}, (2) \cite{Gaspar20}, (3) \cite{nielsen16}, (4) \cite{beta_pic_dust}, (5) \cite{bpic_almaco}, (6) \cite{bpic2006}, (7) \cite{131488_alma}, (8) \cite{melis13}, (9) \cite{Saffe2021}, (10) \cite{song12}, (11) \cite{peacut16}, (12) \cite{Hales2022}, (13) \cite{blanch15}, (14)  \cite{32297_ALMA}, (15) \cite{Debes09}, (16)\cite{disk_gassample}, (17) Matr\`a et al. private communication, (18) This work
     \end{minipage}
     \label{tab:star_params}
 \end{table*}

\section{Disk Vertical Structure}\label{sec:Results}

The non-detection of CO in absorption towards HD 32297 is surprising given that the inclination of the dust disk is well measured to be nearly edge-on and HD 32297 has one of the largest CO masses of debris disks, three orders of magnitude larger than that of $\beta$ Pic (see Table \ref{tab:star_params}). The inclination measured from the GPI scattered light image is 88.88$^{+0.01}_{-0.02}$ degrees \citep{Gaspar20} and the ALMA dust continuum image is 88.4 $\pm0.1$ degrees (Matr\`a et al. private communication). HD 32297 is more edge-on and has a higher CO gas mass than $\beta$ Pic (see Table \ref{tab:star_params}), which has CO detected in absorption in M-band \citep{bpicabsorp}. One possibility for the non-detection of CO around HD 32297 with iSHELL is that the scale height of the CO is small, resulting in a low number of molecules along the line of sight to the star.

\subsection{CO Column Density Estimation}

We estimated the scale height upper limit for HD 32297 using the measured column density upper limit and assuming that the gas in the disk is in hydrostatic equilibrium. In hydrostatic equilibrium, the gas number density can be written as
\begin{equation}
    n=\frac{\Sigma}{\sqrt{2\pi}H}e^{\left({\frac{-z^2}{2H^2}}\right)}
\end{equation}
where $z$ is the distance above the disk midplane, $\Sigma$ is the surface density profile, and $H$ is the scale height and in hydrostatic equilibrium is defined as 
\begin{equation}
    H=\left(\frac{kT_HR^3}{\mu GM_{*}}\right)^{1/2}.
\end{equation}

 Here, $R$ is the distance from the star, $\mu$ is the mean molecular weight, $k$ is the Boltzmann constant, $T_H$ is the gas temperature that corresponds to the CO scale height, $G$ is the gravitational constant, and $M_*$ is the stellar mass. $T_H$ does not directly represent the kinetic temperature of the gas, but is rather a parameterization of the CO scale height, as it represents the temperature required for CO to have a certain scale height above the midplane. For the calculation of $T_H$, we assumed a mean molecular weight of $\mu=14$ $M_H$, where $M_H$ is the mass of the hydrogen atom. We chose this value of $\mu$ as it corresponds to a disk dominated by atomic carbon and oxygen, photodissociation products of CO, and is commonly used to describe the mean molecular weight of gas in debris disks (e.g. \citealt{bpic_almaco}, \citealt{kral17}). 
 
 For  the $\beta$ Pic debris disk, atomic carbon was found to be $\sim70$ times more abundant than CO with HST/STIS \citep{Roberge_2000} and oxygen $\sim6-13$ times with FUSE \citep{Roberge2006}. \cite{cataldi18} found that CI emission seen with ALMA has a similar spatial distribution of CO in the $\beta$ Pic disk, suggesting that the atomic and molecular gas is likely well mixed. We also tested other values of mean molecular weight of $\mu=2.35$ $M_H$, corresponding to primordial gas in a protoplanetary disk \citep{kimura16}, as well as $\mu$ equal to the mass of the CO molecule. We discuss the effects of assuming different values of $\mu$ in Section \ref{sec:Co_scale_est}. 
 
The assumption of hydrostatic equilibrium may not be correct because, as suggested by \citep{marino22}, the scale height might not reflect hydrostatic equilibrium, but rather strong photodissociation of the upper layer and weak vertical mixing. We choose to assume hydrostatic equilibrium to be able to compare with previous studies in the literature that also assume hydrostatic equilibrium for gas in debris disks (e.g. \citealt{chen04,hughes08,bpic_almaco,cataldi18,Hales2022}). If the gas in these disks is not in hydrostatic equilibrium, then our scale height estimates may not be accurate, which is a potential limitation to this method.   
 
\begin{figure*}[!htbp]
    \centering
    \includegraphics[scale=0.75]{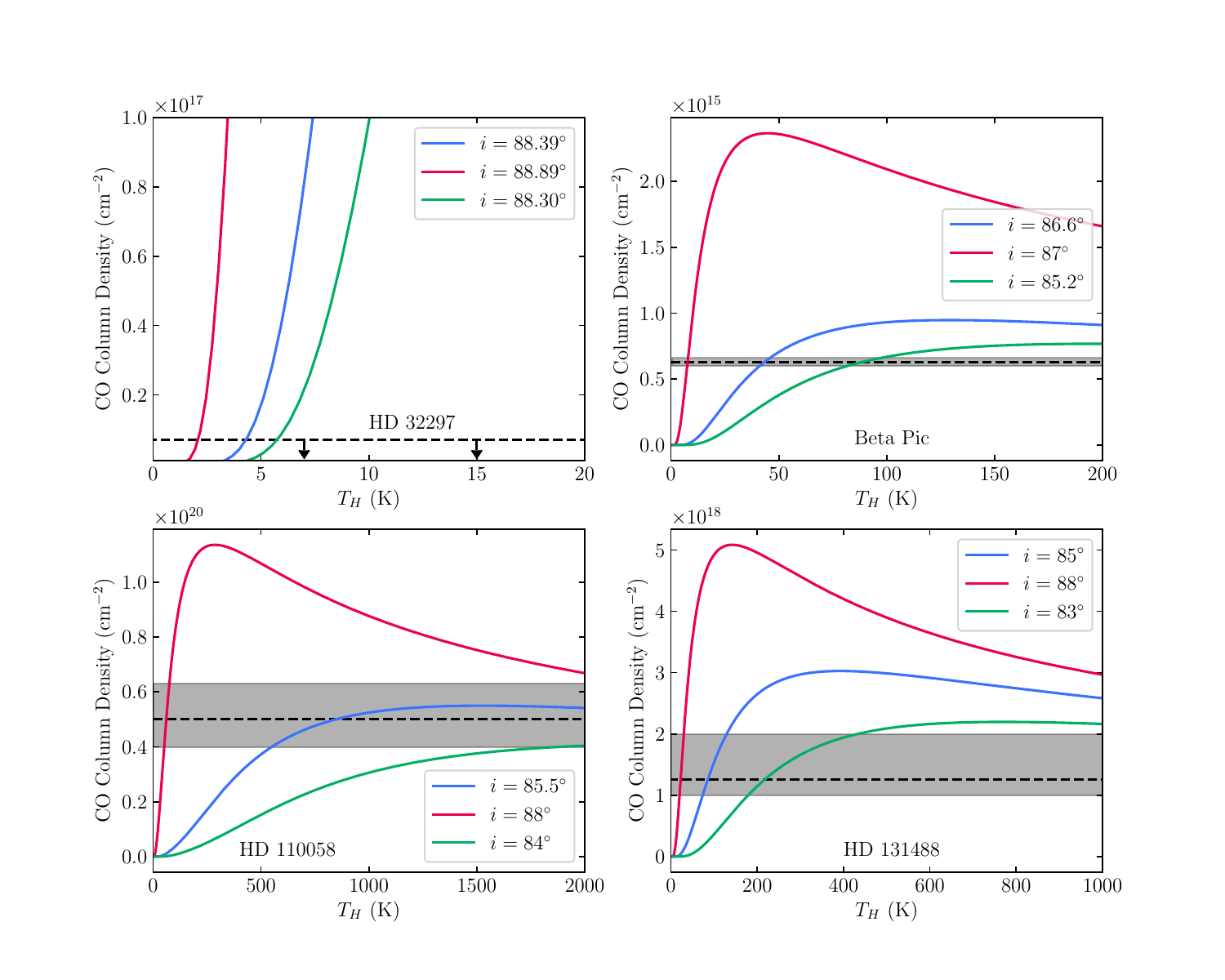}
    \singlespace\caption{CO Column density as a function of scale height temperature ($T_H$) for the four systems assuming that the CO is well mixed with C and O so the gas has $\mu$=14$M_H$. The blue line shows the column density for the best fit disk parameters from Tables 1 and 2, while the orange and green lines show the lower and upper bounds of the CO column density estimates using the uncertainties of the disk properties. We list the inclinations corresponding to each curve as we find that the uncertainty in disk inclination is the dominant source of uncertainty in the estimation of the CO column density. The horizontal black dashed lines show the CO column density upper limit for HD 32297 and the measured \ce{CO} column densities for $\beta$ Pic, HD 131488. and HD 110058. The gray shaded area shows the uncertainty in the column density measurements. We determine $T_H$ for each of these four systems by finding the intersection point of the blue curves and black dashed lines, and the uncertainty in $T_H$ is found by finding the intersection of the measured column densities with the green and orange curves. The estimated $T_H$ values are shown in Table \ref{tab:temp}. For HD 32297, the model column densities reach up to $1\times10^{18}$ cm$^{-2}$, but we only show up to $1\times10^{17}$ cm$^{-2}$ to more clearly show the intersection of the column density upper limit.}
    \label{fig:columnden}
\end{figure*}

\begin{figure}[!htbp]
    \hspace*{-0.7cm}
    \centering
    \includegraphics[scale=0.6]{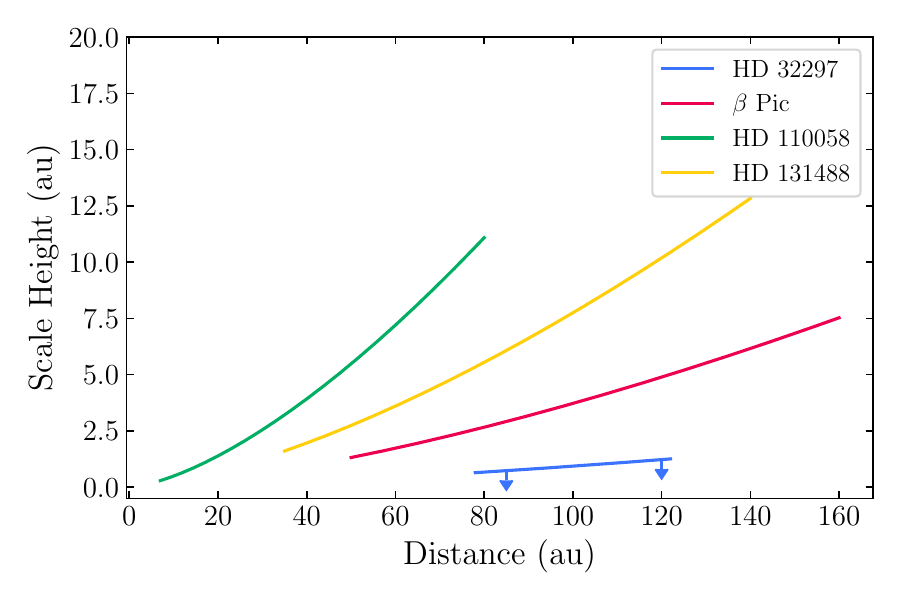}
    \caption{Estimated \ce{CO} scale height as a function of distance from the central star for each of the four systems. The CO scale height shown here for HD 32297 is an upper limit. The scale heights shown here do not depend on the assumption of mean molecular weight, because all combinations of $\mu$ and $T_H$ shown in Table \ref{tab:temp} give the same scale height.}
    \label{fig:scale}
\end{figure}

 We assumed a surface density profile of 
\begin{equation}
    \Sigma=\Sigma_0R^{-3/2}
\end{equation}
as in \cite{solarneb} and \cite{49cetiALMA}. We solved for $\Sigma_0$ by integrating the surface density profile across the radial extent of the disk and equating it to the CO mass measured with ALMA shown in Table \ref{tab:disk_prop}. Once $\Sigma_0$ and the inclination of the disk from the line of sight ($i$) are known, we estimated the column density of the gas using 
\begin{equation} \label{eq:col} \centering
    N=\frac{1}{m_{\ce{CO}}\cos{i}\sqrt{2\pi}}\int_{R_{in}}^{R_{out}} \frac{\Sigma}{H}e^{-sin^2(i)R^2/2H^2} dR
\end{equation}
where $N$ is the column density, $m_{CO}$ is the mass of the CO molecule, $R_{in}$ is the inner disk radius, and $R_{out}$ is the outer radius of the disk as in \cite{chen04}. We estimated the column density as a function of $T_H$ for each of the four systems: HD 32297, $\beta$ Pic, HD 110058, and HD 131488 by numerically integrating Equation \ref{eq:col} across a range of $T_H$ values using the disk parameters listed in Tables 1 and 2. We solved for $T_H$ in $\beta$ Pic, HD 110058 and  HD 131488, and the upper limit for $T_H$ in HD 32297 by setting the predicted column density as a function of $T_H$ equal to the measured column densities of the three targets with CO absorption detected and the column density upper limit of HD 32297 (see Figure \ref{fig:columnden}). The values we estimated for $T_H$ for each of these 4 systems were then used to determine the CO scale height as a function of distance from the star using Equation 4. These scale heights are shown in Figure \ref{fig:scale}. For HD 32297, this method of using the total CO mass from C$^{18}$O emission, the disk inclinations measured from scattered light and ALMA images, and assuming the gas is in hydrostatic equilibrium, constrains the scale height of CO to be $<$2 au across the disk radial extent. In this model, if the CO scale height was larger than 2 au, we would have detected CO absorption in our iSHELL spectrum, based on the column density upper limit. Even though our upper limit is $\sim3$ times the CO column density detected in $\beta$ Pic, we are able to constrain the CO scale height from our iSHELL upper limit under the assumptions listed above. This is because the HD 32297 disk is more edge on by $\sim2$ degrees and the CO mass is three orders of magnitude greater for HD 32297 than $\beta$ Pic, so the non-detection of CO absorption constrains the scale height.

A possible complication of this method is the optical depth of the CO lines used to compute the CO mass of these targets. If the CO lines are optically thick, this would make the CO masses derived from ALMA that we used in the calculation of column density and scale height lower CO mass limits. To overcome this, we used CO masses for HD 131488 and HD 32297 that are derived from optically thin C$^{18}$O emission from \cite{131488_alma,moor19}. For $\beta$ Pic, the CO emission seen with ALMA was found to be optically thin \citep{bpic_almaco}, enabling an accurate measurement of the CO mass. For HD 110058, the CO mass is derived from forward modeling $^{12}$CO and $^{13}$CO lines \citep{Hales2022}, so the optical depth of the lines is taken into account in the uncertainty on the total disk CO mass. We propagated this uncertainty in CO mass in the uncertainty in the CO scale height. The CO mass for each of these disks is shown in Table \ref{tab:star_params}.

Another possible complication of this method, in the case of $\beta$ Pic, is the discrepancy between the observed column densities of CO in $\beta$ Pic at IR and UV wavelengths. The CO column density from the IR \citep{bpicabsorp} is $\sim$3 times larger than that from the UV \citep{Roberge_2000}. A possible explanation for this discrepancy is that some of the UV absorption lines are saturated, providing an underestimate of the total CO column density \citep{bpicabsorp}. We included this discrepancy of CO column density towards $\beta$ Pic as a source of uncertainty in the CO scale height calculation.

We used direct measurements of the inner and outer radius of the gas for $\beta$ Pic, HD 110058, and HD 131488 from ALMA CO observations \citep{bpic_almaco,moor19,Hales2022} to estimate the column density along the line of sight to the star. For HD 32297, we used the inner and outer radii of the dust since the radial distribution of the gas and the dust was found to be similar and no direct measurement of the radii of the gas was made as the gas was found to likely have sub-Keplerian velocities \citep{32297_ALMA}. $\beta$ Pic and HD 32297 have inclinations measured from both scattered light images and ALMA continuum images in the literature \citep{Blanchaer15,beta_pic_dust,Gaspar20}. For the system (HD 110058) with a precise inclination measured from ALMA CO observations, we used this inclination. Otherwise, we used the inclinations from the ALMA continuum images of these sources, but determined the uncertainties in inclination from the difference between the scattered light and ALMA inclination measurements. The inclinations measured from ALMA and GPI for these systems are consistent within $\pm1.5$\textdegree and 2-3$\sigma$ of each other (See Table \ref{tab:star_params}). We include this discrepancy in inclination of these disks from scattered light and ALMA images in the uncertainty propagation for the estimated CO scale height, so this inclination uncertainty is represented in the estimated CO scale height uncertainty. HD 110058 and HD 131488 do not have inclinations measured from scattered light presented in the literature, so we used ALMA measured inclinations. For HD 110058, we determined the outer radius using ALMA archival data, as this value is not currently present in the literature (see Appendix A).

\begin{table*}[!htbp]

    \centering
    \caption{Disk properties used in column density calculations.}
    \begin{tabular}{c c c c c c }
    \hline
        System & Measured CO $R_{in}$ (au) &Measured CO $R_{out}$ (au) & CO Column Density ($cm^{-2}$) &$R_{out}$ dust & References \\
    \hline
         HD 32297& 78.5$\pm8.1$* & 122$\pm$3* & $<$6$\times 10^{15}$ &122  & (1),(2), (11) \\
        $\beta$ Pic &50$\pm$5 & 160$\pm$5 & 6.3$\pm0.3$ $\times10^{14}$ & 150& (3),(4),(5),(6)  \\
      HD 131488& 35$\pm$11 & 140$\pm$11 &1.26$^{+0.73}_{-0.26}$ $\times 10^{18}$& 115 & (7), (8), (9)\\
        HD 110058& 7.4$^{+2.2}_{-7.3}$&80$\pm$12& 5.01$^{+1.30}_{-1.03}$ $\times 10^{19}$& 70& (1),(9), (10), (11) \\
    \hline
    \end{tabular}
    \begin{minipage}{16.5cm}
    \vspace{0.1cm}
    
     \textbf{Notes:} *CO inner and outer radii not available in literature, so dust inner and outer radii is used instead. \textbf{References:} (1) this work (2) \cite{32297_ALMA}, (3) \cite{beta_pic_dust}, (4) \cite{Dent14}, (5) \cite{bpic_almaco}, (6) \cite{Roberge_2000}, (7) \cite{131488_alma}, (8) \cite{disk_gassample}, (9) Brennan et al. (submitted.), (10) \cite{Hales2022},
     (11) \cite{sifry16}
    \end{minipage}
    \label{tab:disk_prop}
\end{table*}

\subsection{Estimated CO Scale Height}\label{sec:Co_scale_est}

We find that HD 131488 has the largest \ce{CO} scale height, followed by HD 110058, $\beta$ Pic, and HD 32297. For $\beta$ Pic, the derived $T_H$ values correspond to temperature that is significantly larger (the error bars do not overlap) than the excitation temperature measured from UV and IR absorption line spectroscopy (see Table \ref{tab:temp}). This large $T_H$ value suggests that the $\beta$ Pic scale height cannot be solely explained by the measured gas temperature. $\beta$ Pic is the only one of these systems to have a CO scale height measured with ALMA. Figure 3 in \cite{bpic_almaco} shows the CO scale height of $\beta$ Pic to be between $\sim5-12$ au at the outer edge of the disk. Using our method and the assumptions listed above, we estimate a CO scale height of 7$^{+4}_{-1}$ au at the outer edge of the $\beta$ Pic disk, which is consistent with the CO scale height measured in \cite{bpic_almaco}, giving some confidence that this method of CO scale height estimation is reasonable. For HD 110058 and HD 131488, the CO excitation temperature is within the uncertainties of the estimated value of $T_H$ because the uncertainty in disk inclination creates a large uncertainty in $T_H$ (see Figure 4). HD 110058 and HD 131488 have excitation temperatures measured from ro-vibronic bands of CO absorption with HST STIS (Brennan et al. submitted). Therefore, we cannot definitively conclude that $T_H$ is greater than the excitation temperature for HD 110058 using the assumption of a mean molecular weight of $\mu=14 M_H$ and with the current measurements of the disk geometry. If the inclination of the HD 110058 disk is less than $\sim$87 degrees, then $T_H$ would be significantly larger than the measured excitation temperature.   

We test three different values of mean molecular weight ($\mu$) in estimating the CO scale heights and $T_H$ values of these systems. They are: $\mu$ equal to the mass of the CO molecule,  $\mu$ corresponding to a disk dominated by atomic C and O ($\mu$=14 $M_H$), and $\mu$ for a primordial gas disk dominated by \ce{H2} where $\mu$=2.35 $M_H$. The assumption of $\mu$ does not change the estimated CO scale height because, by numerically integrating Equation 6 and equating it to the observed column density of CO, we solve for the entire quantity $\left(\frac{kT_H}{\mu G M_*}\right)$ from Equation 4 all at once. Thus, the estimated scale height as a function of radius does not depend on our assumption of $\mu$ (all combinations of $T_H$ and $\mu$ shown in Table \ref{tab:temp} for the same system give the same scale height). However, the value of $T_H$ that we estimate does depend on the value of $\mu$ we assume. A larger value of $\mu$ corresponds to a larger $T_H$ value for a given scale height. The different $T_H$ values for different assumptions of $\mu$ are shown in Table \ref{tab:temp}.

\subsection{ALMA Continuum}\label{sec:alma_cont}

 We compared the estimated CO and millimeter dust scale heights in these four systems to look for similarities between the gas and dust scale heights. We used the literature values of the dust scale height of $\beta$ Pic \citep{beta_pic_dust}, HD 110058 \citep{Hales2022}, and HD 32297 (Matr\`a et al. private communication, Progam: 2021.1.01477.S) from ALMA continuum images. We estimated dust scale heights from the publically available visibility data of HD 131488 in band 6 \citep{131488_alma}. To estimate the dust scale height, we modeled the visibilites using a similar method to \cite{beta_pic_dust} where we used RADMC-3D \citep{Dullemond12} to compute the radiative transfer at 1.33 mm. We did this using the code Modelling Interferometric Array Observations\footnote{\url{ https://github.com/dlmatra/miao}} (MIAO, Luca Matr\`a). We assumed a dust radial temperature profile of blackbody grains around HD 131488 with a stellar luminosity of $L_*=15.5 L_{\odot}$ \citep{melis13}. We assumed the same dust grain properties as in \cite{beta_pic_dust}. We assumed a dust density distribution that is axisymmetric, Gaussian in both radial and vertical profiles, and has a constant aspect ratio $h=H/r$ where $H$ is the scale height and $r$ is the disk radius. This density distribution is described by the equation 
\begin{equation}
    \large\rho=\Sigma_0e^{\frac{-(r-R)^2}{2\sigma^2}}\frac{e^{-\frac{z^2}{2(hr)^2}}}{\sqrt{2\pi}hr},
\end{equation}
where $r$ is the radius, $z$ is the height above the midplane in cylindrical coordinates, $h$ is the aspect ratio ($H/r$), $R$ is the central Gaussian disk radius, and $\sigma$ is the radial Gaussian width of the disk and the radial FWHM of the disk is computed from $\sigma$ and is represented by $\Delta R$, and $\Sigma_0$ is a normalization factor proportional to the mass of the disk. In the visibility fitting, $h$, $R$, $\Delta R$, $\Sigma_0$, as well as disk inclination and position angle are all left as free parameters. Once we created the image of the model belt with RADMC-3D, we produced model visibilities by first shifting the belt by R.A. and dec offsets that we leave as free parameters in the fit. We then multiplied the image by the primary beam of the observations and Fourier transformed the model image to produce complex model visibilities that we evaluated at the $u-v$ locations sampled by the observations, as described in \cite{marino16}. The model visibilities were then fit to the data with a Monte Carlo Markov Chain (MCMC) method implemented through the EMCEE package \citep{Foreman13}. The MCMC was run with 90 walkers for 4000 steps and we used the EMCEE package to infer the posterior distribution of each parameter. For each parameter, we assumed uniform priors listed in Table \ref{tab:vis_model}. To determine the uncertainties ($\sigma$) on the visibility data points, we used the same method as \cite{beta_pic_dust}, which used the visibility weights ($w$) delivered by ALMA where $\sigma_{u_i-v_i}=1/\sqrt{w_{u_i-v_i}}$. We left the scaling of the weights to be a free parameter to ensure that the visibility uncertainties are correct in the absolute sense (e.g. \citealt{Marino18,beta_pic_dust,Matra20}).

\begin{figure*}[!htpb]
    \centering
    
    \includegraphics[scale=0.28]{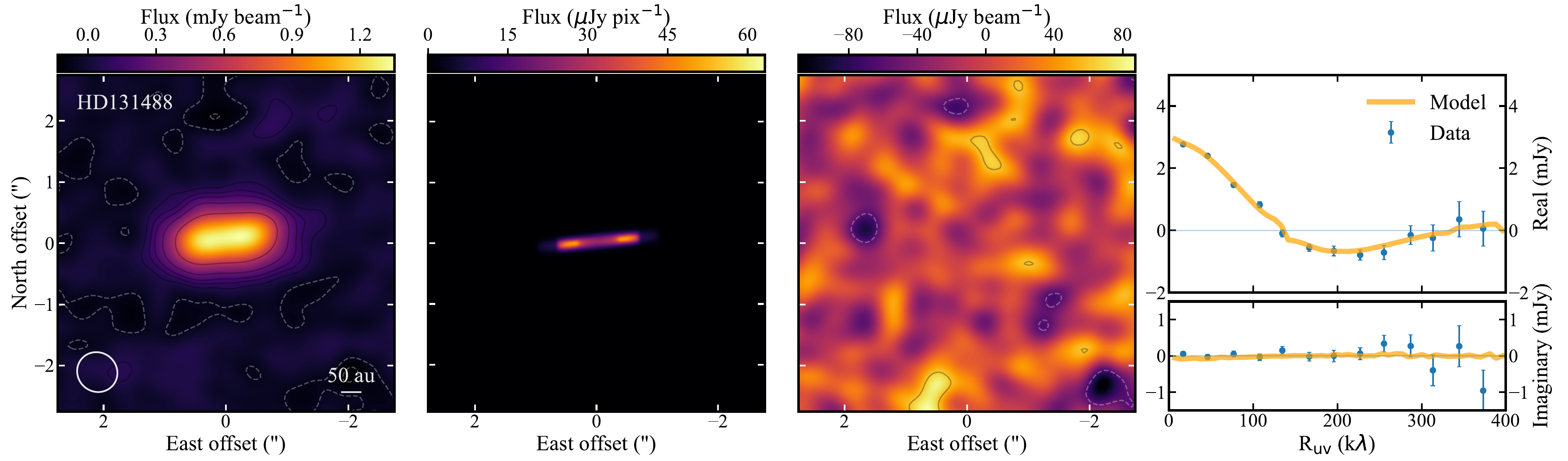}
    \caption{Left: ALMA 1.33 mm continuum image of HD 131488. The contours show $\pm$2,4,6, and 10 times the image rms. The dashed lines show the negative contours while the solid lines show the positive contours. The circle shows the size of the primary beam. Second from left: Best fit 1.33 mm model continuum image for the HD 131488 disk. Third: Residual image of the HD 131488 image and the best fit model image, after convolving with the primary beam. The solid lines show positive contours that are 2 times the image rms and the dashed lines show the negative contours that are 2 times the rms. Right: Deprojected real and imaginary visibilites (blue) with the best fit model visibility (yellow). The deprojection is done using the best fit model geometry as well as averaging the visibilities in bins of 30 k$\lambda$.  }
    \label{fig:HD131488_vis}
\end{figure*}

\begin{figure*}[!htbp]
    \centering
    \includegraphics[scale=0.52]{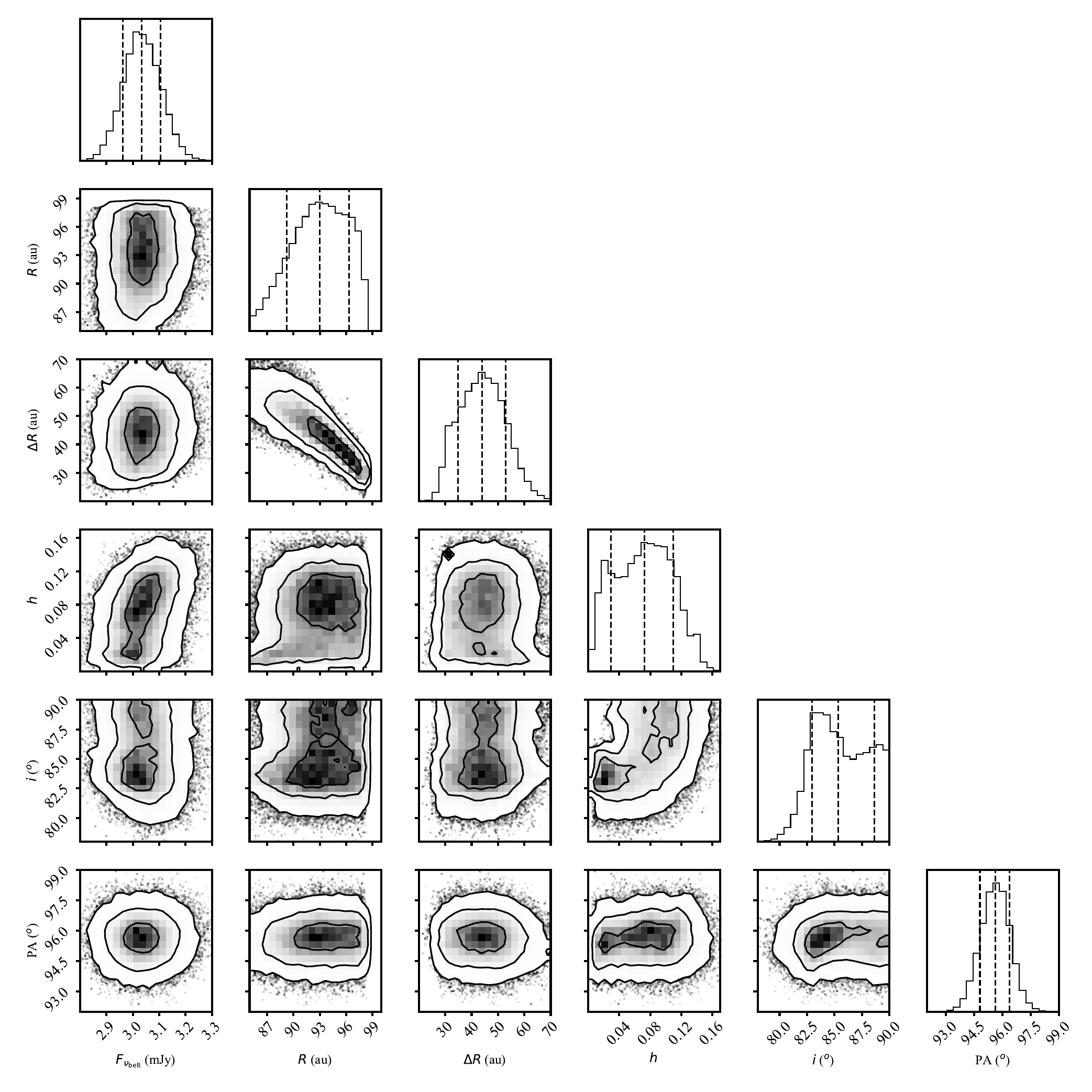}
    \caption{Corner plot showing the posterior distributions of the model parameters from the visibility modeling of HD 131488. The corner plot appears to show a correlation between the aspect ratio ($h$) and the inclination ($i$). The dashed lines show the 50th, 16th, and 84th quantiles. }
    \label{fig:corner_plot}
\end{figure*}

The best fit parameters from the visibility modeling of HD 131488, which are the 50$^{+34}_{-34}$th percentiles of their marginalized posterior distributions, are shown in Table \ref{tab:vis_model}. The HD 131488 continuum image and the best fit model is shown in Figure \ref{fig:HD131488_vis}. The posterior distributions of the model parameters are shown in Figure \ref{fig:corner_plot}. As shown in the corner plot, there is a degeneracy between the disk inclination and the aspect ratio $h$. The best fit aspect ratio is $h=0.08^{+0.04}_{-0.04}$ which is consistent with an upper limit on the scale height at the lower end of the uncertainty (corresponding to lower inclinations). If the disk inclination, however, is $\gtrsim$84 degrees, then the aspect of the disk would be $h\gtrsim$0.08. Higher angular resolution ALMA images could potentially ameliorate this issue and provide a better measurement of the millimeter dust scale height.

If the disk is inclined such that the vertical structure is resolved, we can estimate the millimeter dust scale height at the outer edge of the disk for comparison with the estimated CO scale height at the disk outer edge. As done in \cite{Hales2022}, we calculate the outer radius of the disk as $R_{out}=R+\Delta R/2$ which gives $R_{out}=115^{+8}_{-9}$ au. The dust scale height at the outer disk radius, from our best fit $h$ and assuming an inclination $>$84$^{\circ}$, is $H=9^{+5}_{-4}$ au. This calculation does not consider that the scale height may be unresolved and thus an upper limit. Instead, it assumes the disk has an inclination $>$84$^{\circ}$ and thus $h\ge0.08$. However, this might not actually reflect the true scale height of the disk because the visibility modeling found that the scale height is still consistent with an upper limit, depending on the inclination. 

\begin{table}[]
    \centering
    \caption{Best Fit Model Parameters}
    
    \begin{tabular}{c c c c }
    \hline
         Parameter& Unit & Best fit Value& Prior range  \\
    \hline
         R & au  &93$^{+4}_{-3}$&[45, 110] \\
         $\Delta$R & au &43$^{+9}_{-12}$&[8, 100] \\ 
         $F_{belt}$& mJy &3.0$^{+0.1}_{-0.1}$&[0.1, 5.0] \\
         inclination & degrees&85$^{+3}_{-2}$&[70, 90] \\ 
         Position Angle & degrees &96$^{+1}_{-1}$&[0, 180] \\
         $h$& N/A &$0.08^{+0.04}_{-0.04}$&[0.005, 0.2] \\
    \hline
    \end{tabular}
    
    \label{tab:vis_model}
\end{table}

\cite{beta_pic_dust} measured the average scale height of $\beta$ Pic across the disk midplane to be 8.2 au  with a single vertical Gaussian fit. \cite{beta_pic_dust} found that a double Gaussian is a better fit to the vertical structure of the $\beta$ Pic disk than a single Gaussian profile, but we used the measurement from the single Gaussian profile to be consistent with the method we used to measure the scale heights for HD 32297, HD 110058, and HD 131488. The vertical Gaussian fit from the visibility modeling of the the HD 32297 ALMA continuum data, that has an angular resolution of 0.05'', gave an aspect ratio of $H/r=0.015\pm0.002$. At the dust outer radius of 122 au, this gives a scale height of  1.83$\pm$0.2 au (Matr\`a et al. private communication). Forward modeling of the visibilities for HD 110058 yielded an aspect ratio of $h=0.17^{+0.05}_{-0.09}$ \citep{Hales2022} with a degeneracy with the disk inclination. However, modeling of the gas as well as well as scattered light images suggested that the disk has an inclination $>80^{\circ}$ \citep{Esposito20,Hales2022}. With this constraint on the inclination, \cite{Hales2022} found that the disk is vertically thick with a vertical aspect ratio of $h=0.13-0.28$. The angular resolution of the scattered light observations are better than that of the ALMA continuum and the inclination measured from forward modeling the CO is also more precise than that of the ALMA continuum, suggesting that the disk does have $i>80^{\circ}$. Because of this, we used the aspect ratio of $h=0.13-0.28$ from \cite{Hales2022} for HD 110058.

\subsection{Trends between millimeter/micron dust and CO scale heights }
\begin{figure}[!htbp]
    \centering
    \includegraphics[scale=0.55]{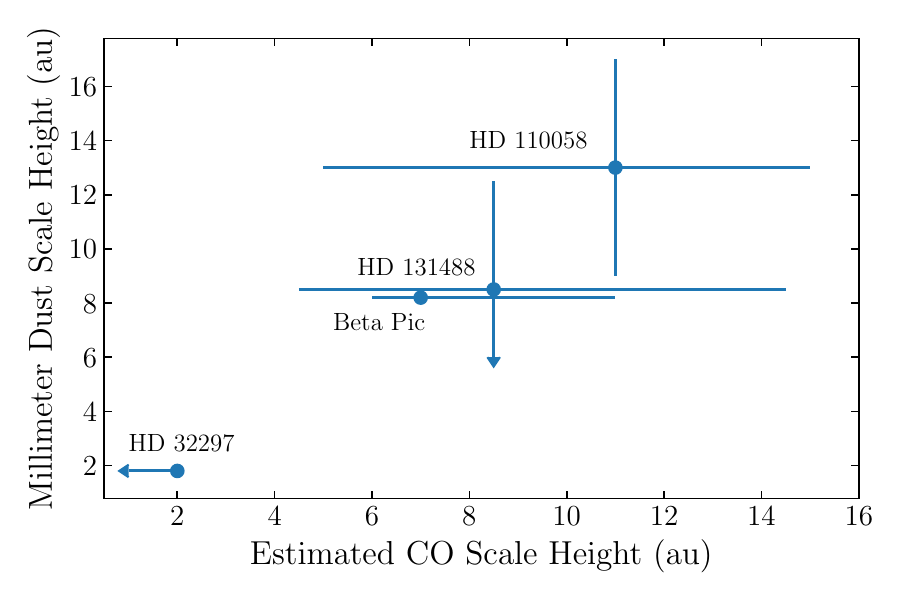}
    \caption{Dust scale height measured from ALMA versus \ce{CO} scale height we derive at the dust outer radius for each of these systems. The \ce{CO} scale height shown for HD 32297 is an upper limit. Both $\beta$ Pic and HD 32297 have error bars on their dust scale height, their uncertainties are too small to be seen in this figure. The millimeter dust scale height of HD 131488 is also consistent with an upper limit on the lower end of the uncertainty. }
    \label{fig:corr}
\end{figure}

 We search for a trend between the estimated \ce{CO} scale heights and the ALMA dust scale heights for these four systems. Figure \ref{fig:corr} shows the ALMA dust scale height plotted as a function of the \ce{CO} scale height we estimated from absorption line spectroscopy. To facilitate comparison between the CO and the millimeter dust, the estimated CO scale height shown in Figure \ref{fig:corr} is calculated at the dust outer radius. We find that for each of the four systems, the estimated CO scale height and the dust scale height are consistent within the uncertainties (see Table \ref{tab:scale_heights}), suggesting that the dust and CO have similar vertical distributions in each of the systems. The uncertainties in the \ce{CO} scale height are determined by propagating the uncertainties in inclination and radial distribution of each disk through the calculation of the gas scale height.

To quantify any potential correlation between the estimated CO scale height and the millimeter dust scale height, we performed a Monte Carlo simulation of 10,000 iterations to sample the uncertainty range of the dust and estimated CO scale heights of each system. At each iteration, we sampled a Gaussian distribution of the CO and dust scale heights for each of the four systems where the center of the Gaussian is the best fit scale height and the Gaussian $\sigma$ is the uncertainty (as shown in Figure \ref{fig:corr}). We then calculated a Pearson's $r$-value and a $p$-value at each iteration to check for a positive correlation. From this Monte Carlo simulation, we produced a distribution of 10,000 $r$ and $p$-values. The probability distribution of the $r$ and $p$-values are shown in Figure \ref{fig:r-values}. These probability distributions were calculated such that the integral of all of the bins is equivalent to 1.
 
We determine a Pearson's $r$-value from this Monte Carlo simulation to be $r=0.80\pm0.25$, where 0.80 is the median of the distribution and 0.25 is 1$\sigma$. With the distribution of 10,000 $r$-values, we estimate a standard error on the $r$-value of 0.0025. This $r$-value suggests that there is a positive linear correlation between the estimated CO scale heights and the millimeter dust scale heights of these four systems. We are able to rule out Pearson's $r$ of $r=0$ at the 3$\sigma$ level. We found a $p$-value of $p=0.08\pm0.11$, where 0.08 is the median and 0.11 is the 1$\sigma$ of the distribution. The probability distribution of the $p$-values peaks below 0.05, however, the distribution of $p$-values makes it so we are unable to definitively conclude that the correlation is statistically significant at the $5\%$ level. Because of this, we conclude that there is a potential correlation between the estimated CO scale heights and the millimeter dust scale heights in these four systems since we cannot say for certain that the correlation is statistically significant. We also call this a potential correlation because the scale height of the HD 131488 millimeter dust is consistent with an upper limit from the forward modeling. This potential correlation is also based on the assumption that the CO is in hydrostatic equilibrium in these four systems, which as discussed above, may not be a correct assumption.

\begin{figure}[!htbp]
    \centering
    \includegraphics[scale=0.55]{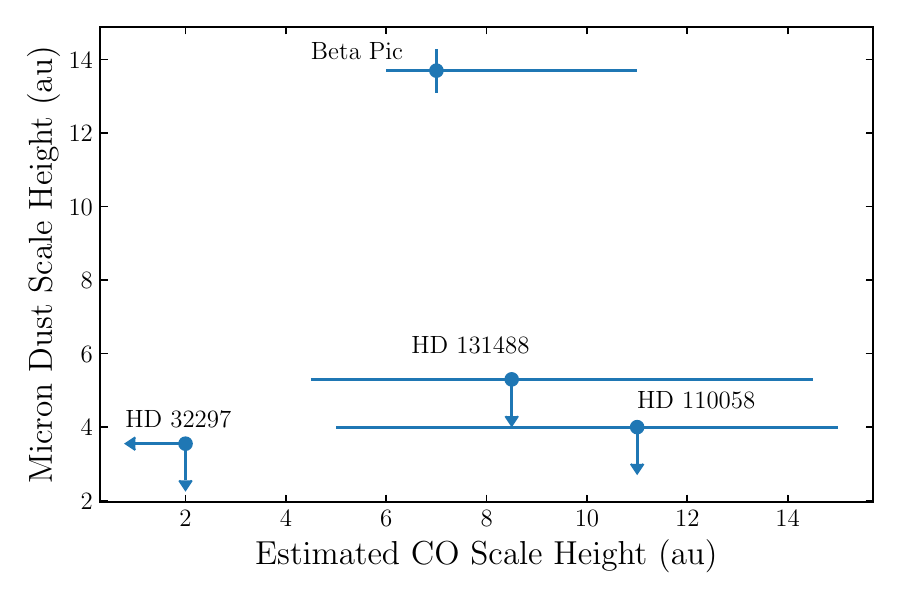}
    \caption{Dust scale height from scattered light images versus \ce{CO} scale height we derive at the dust outer radius for each of these systems. The \ce{CO} scale height shown for HD 32297 is an upper limit. The scattered light scale heights for HD 32297, HD 110058, and HD 131488 are upper limits.}
    \label{fig:micron_scat}
\end{figure}

\begin{figure*}[!htbp]
     \centering
     \includegraphics[scale=0.8]{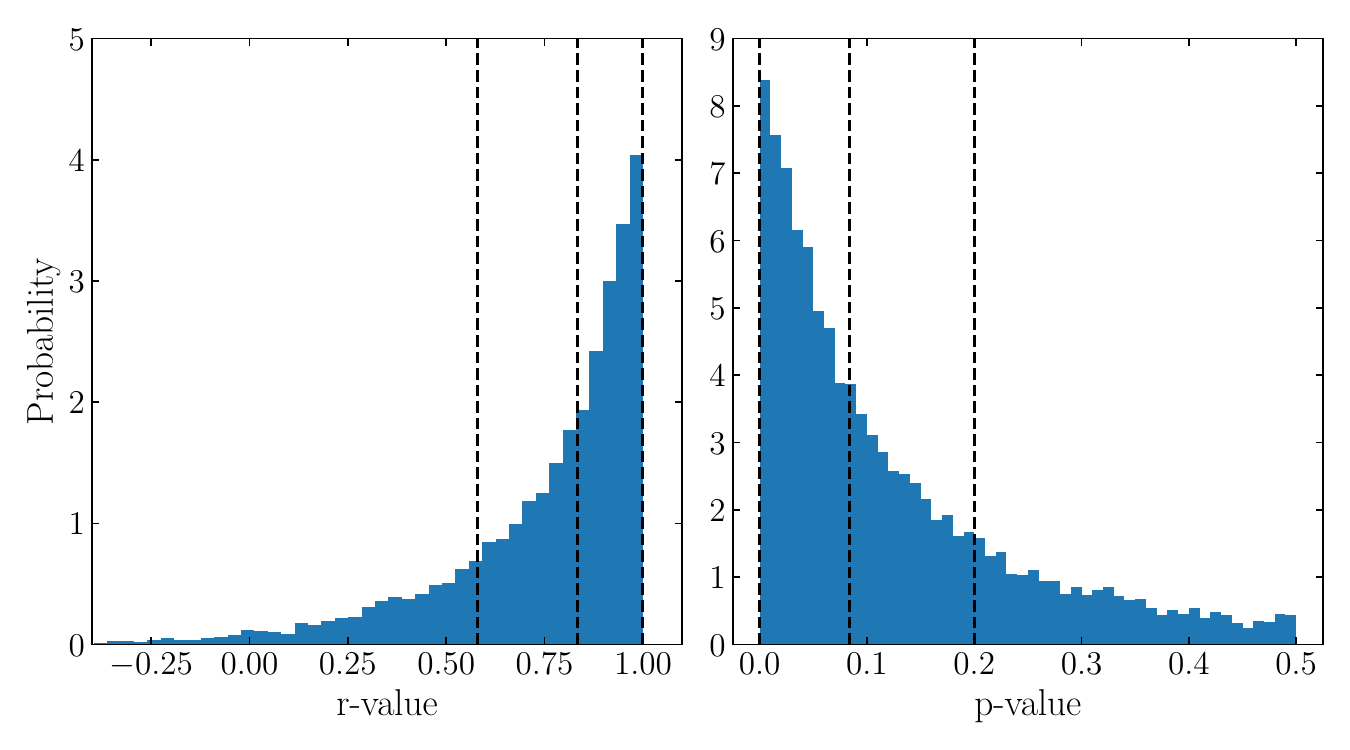}
     \caption{Left: Probability distribution of Pearson's $r$-values from the Monte Carlo simulation to search for a correlation between the estimated CO scale height and millimeter dust scale height. Right: Corresponding $p$-value distribution of the same test. The vertical black dashed lines show the median $r$ and $p$-values$\pm 1\sigma$ from the distribution. Because the median$\pm$1$\sigma$ goes above $r$=1 and below $p$=0, we set the max and min of the black lines to be at $r$=1 and $p$=0 respectively.}
     \label{fig:r-values}
 \end{figure*}

To trace the vertical distribution of the micron-sized dust grains in these disks, we also look for trends in scale height of the scattered light images. A value of the scattered light scale height is reported for $\beta$ Pic in \cite{Blanchaer15}. \cite{Gaspar20} report a scattered light scale height for HD 32297, however they do not definitively conclude that they resolved the HD 32297 disk vertically with GPI, so we treat this scale height as an upper limit. From our GPI H-band image, we put an upper limit of the scale height of HD 131488 in scattered light to be $<\sim$5-6 au. Similarly, HD 110058 is not vertically resolved in H-band with SPHERE and has a scale height upper limit of $<\sim$4 au in scattered light \citep{Kasper15}. The micron-sized dust scale height is plotted as a function of estimated CO scale height in Figure \ref{fig:micron_scat} (see Table \ref{tab:scale_heights} for values). We do not see the same trend as with the millimeter and estimated CO scale heights, however, three of the scattered light scale heights are upper limits, so it is difficult to identify any trends.

\begin{table*}[!htbp]

    \centering
    \caption{Excitation temperatures and $T_H$ values using different mean molecular weights of the gas.}
    \begin{tabular}{c c c c c c c }
    \hline
        Source & $T_{ex}$ UV & $T_{ex}$ IR & $T_H$($\mu$=C+O)*& $T_H$($\mu$=CO)& $T_H$($\mu$=\ce{H2}) & References \\
    \hline
    HD 32297& NA & NA &$<5$ K &$<10$ K & $<2$ K& (1)  \\
        $\beta$ Pic &  15.8$\pm0.6$ K &15$\pm$2 K &40$^{+50}_{-20}$ K &80$^{+61}_{-12}$ K & 7$^{+8}_{-3}$ K &(1),(3),(4)  \\
        HD 110058& 72$^{+3}_{-3}$ K  &NA&880$^{+1000}_{-820}$ K & 1700$^{+1500}_{-500}$ K & 150$^{+150}_{-130}$ K&(1),(2)  \\
        HD 131488 & 45$^{+8}_{-8}$ K & NA& 100$^{+300}_{-70}$ K & 200$^{+600}_{-140}$ K & 20$^{+80}_{-10}$ K&(1),(2) \\
    \hline
    \end{tabular}
    \begin{minipage}{10.5 cm}
    \vspace{0.1cm}
   
    \textbf{Notes:} *These results for $T_H$ are shown in Figure \ref{fig:columnden}. \textbf{References:} (1) This work, (2) Brennan et al. (submitted), (3) \cite{bpicabsorp}, (4) \cite{Roberge_2000}
     \end{minipage}
    \label{tab:temp}
\end{table*}

\begin{table*}[!htbp]
    \centering
    \caption{Dust and estimated CO scale heights of the four debris disks}
    \begin{tabular}{c c c c c }
    \hline
        Source& ALMA dust scale height& Scattered light scale height& Estimated CO scale height at dust outer radius*
     & References  \\
        \hline
        HD 32297 &1.8$\pm$0.2 & $<$3.5 & $\lesssim 2$& (1),(2),(5) \\
        $\beta$ Pic& 8.2$\pm$0.2 &13.7$^{+0.5}_{-0.6}$ &7$^{+4}_{-1}$& (1),(3),(4) \\
        HD 110058 &13$^{+5}_{-4}$ &$<4$&11$^{+4}_{-6}$& (1),(6) \\
        HD 131488  &9$^{+5}_{-4}$ &$<6$ &8$^{+6}_{-4}$& (1) \\
         \hline
    \end{tabular}
    \begin{minipage}{17 cm}
    \vspace{0.1cm} 

     \textbf{Notes:} All scale heights in this table are in units of au. *Dust outer radii are listed in Table \ref{tab:disk_prop}. \textbf{References:} 1. This work 2. Matr\`a et al. Private Communication 3. \cite{beta_pic_dust} 4. \cite{Blanchaer15} 5. \cite{Gaspar20} 6. \cite{Kasper15}
    \end{minipage}
    \label{tab:scale_heights}
\end{table*}

\section{Discussion}
\subsection{Origin of the Disk Scale Heights}
 The inclination distribution of the bodies in our Kuiper Belt has been used to constrain the dynamical evolution of our solar system. The high inclination of the ``hot" population of Kuiper Belt Objects can be explained by gravitational interactions with a slowly migrating Neptune \citep{nesvorny2015}. More generally, the inclination of planetesimals in a disk can be excited by secular perturbations from planetary mass bodies within the disk (e.g. \citealt{Daswon2011}) as well as scattering from gravitational interactions with planets (e.g. \citealt{Brown2001}). 
 
 For $\beta$ Pic, \cite{beta_pic_dust} suggest that the high inclination of the millimeter-sized dust observed with ALMA could be caused by gravitational interactions with planet-mass bodies within the system. The $\beta$ Pic system has 2 known planets: $\beta$ Pic b \citep{lagrange10} and $\beta$ Pic c \citep{gravitybpicc}. However, \cite{beta_pic_dust} require an additional planet to explain the inclination distribution of the millimeter-sized dust. If both the millimeter-sized dust and CO in debris disks are produced from the same population of planetesimals, then it is possible that the CO scale height in debris disks can also be altered by the presence of planets. The $T_H$ and corresponding CO scale height values that we estimate for $\beta$ Pic and possibly HD 110058 and HD 131488 could thus be due to dynamical excitation, potentially from unseen planets currently below the detection limit of direct imaging, because they might not be explained by the measured gas excitation temperature alone.   

The vertical structure of the micron-sized dust in a debris disk can be additionally impacted by radiation pressure and collisional interactions between dust grains within the disk. The micron-sized dust grains are thus not expected to be vertically flat and are predicted to have a minimum aspect ratio of $h/R=0.04$ \citep{Thebault09}. The millimeter-sized dust grains are expected to remain settled at the midplane in the absence of a perturbing body and thus have an aspect ratio of $h/R<0.04$ \citep{Thebault09}. The scale heights of the millimeter-sized dust particles of the disks that are vertically resolved with ALMA, $\beta$ Pic, HD 110058, and  HD 131488 all have aspect ratios $h/R>0.04$. Therefore, the vertical structure of the millimeter-sized particles in these disks cannot be explained solely by radiation forces and collisions among particles.

\textbf{HD 32297:} The GPI scattered light images of HD 32297 \citep{Gaspar20} do not show evidence of strong disk-planet interactions. \cite{Gaspar20} found a scale height aspect ratio of $h/R\sim0.04$ for the GPI image of HD 32297. This is consistent with the predicted scale height created from radiation forces and collisions between dust particles from \cite{Thebault09}. The scale height aspect ratio of the millimeter-sized dust from ALMA is also relatively small and is $h/R\sim0.015$ ($H$$\sim$1.8 at 120 au) (Matr\`a et al. private communication). Our estimated 3-$\sigma$ CO scale height upper limit of $\lesssim 2$ au is consistent with the dust scale height from ALMA. From the estimated \ce{CO} scale height and $T_H$ upper limits for HD 32297, we similarly find no evidence for perturbations by a planet in the system. The estimated CO scale height can be explained by a gas temperature less than 30 K, which is appropriate at the location of the disk $\sim$80-120 au \citep{32297_ALMA}. Thus, our findings on the vertical structure of HD 32297 are consistent with the GPI H-band image that suggested HD 32297 is a  dynamically ``cold" system.

\textbf{$\beta$ Pic:} The scattered light images of $\beta$ Pic from HST/STIS \citep{Apai15} and VLT/NaCo 
 \citep{Milli14} both show a warp in the inner disk. This warp has been attributed to gravitational interactions with the directly imaged planet $\beta$ Pic b \citep{Nesvold15}. This warp structure was not seen in ALMA continuum images, however the vertical structure of the millimeter-sized dust seen with ALMA could potentially be due to interactions with planet-mass bodies in the system \citep{beta_pic_dust}. We found that the estimated CO scale height is too large to be explained by the measured CO excitation temperature alone (see Table \ref{tab:temp}), and could thus potentially have the same origin proposed for the scale height of the millimeter-sized dust seen with ALMA, if the scale height of CO is at all related to the velocity distribution of the planetesimals.

\textbf{HD 110058:} The scattered light images of HD 110058 show evidence for dynamical interactions in the structure of the disk. Images from both SPHERE and GPI show a counterclockwise warp on both sides of the disk \citep{Kasper15,Esposito20}, similar to the warp seen in the $\beta$ Pic disk in scattered light. HD 110058 does not have any directly imaged companions, but the warp and the extent of the disk in the GPI polarized light image suggests that any potential warp-inducing substellar companion is likely interior to 40 au \citep{Esposito20}.  We, however, cannot conclude that our estimated CO scale height for HD 110058 is greater than the measured excitation temperature due to the large uncertainties on the disk inclination and therefore the CO scale height. A more precise measurement of the disk inclination is required to determine if our estimated CO scale height can be explained solely by the measured excitation temperature.

\textbf{HD 131488:} The estimated $T_H$ value and corresponding CO scale height are within the uncertainties of the measured CO excitation temperature from absorption in the UV with HST (see Table \ref{tab:temp}). Although, if the disk inclination is less than 87$^{\circ}$, then the estimated $T_H$ value would be larger than the measured excitation temperature of the gas. The scale height from the ALMA continuum data is consistent with an upper limit, but could potentially be resolved vertically if the disk inclination is greater than $\sim84^{\circ}$. This uncertainty in the inclination, and thus the dust and gas scale heights could potentially be resolved with higher angular resolution ALMA data, such as in the upcoming ARKS ALMA Large Program (2022.1.00338.L), but this is currently the level of uncertainty we have with the given data.

\subsection{Origin of the CO in these systems}
 The leading explanation for the origin of the gas in debris disks is that the gas is of secondary origin released from collisions between minor bodies or outgassing \citep{marino16,kral2019}. In this case the dust and the gas in debris disks are expected to have similar spatial distributions because they have the same origin from collisions within a population of planetesimals (\citealt{Kospal13},\citealt{hughes18}). Another possibility is that the dust in these systems is secondary while the gas is primordial from the protoplanetary disk. The latter scenario was favored for the HD 21997 debris disk because the dust and the \ce{CO} gas are not co-located \citep{Kospal13}. Similar vertical distribution does not definitively prove that the gas and the dust have the same origin, however, to be consistent with the secondary origin scenario to first order, it is expected that the gas and dust would have similar spatial distributions (e.g. \citealt{hughes18,kral2019}). There are other factors that can influence the scale height of the gas and dust in disks such as turbulence, gas-dust interactions, and photodissociation, so a similar vertical distribution of gas and dust would not definitively prove similar origin.
 
In protoplanetary disks, however, which are also subject to turbulence and gas-dust interactions, the scale height of the millimeter sized dust is typically smaller than that of the more vertically extended gas (e.g. \citealt{Villenave20,olofsson22,Villenave22}). \cite{Villenave20} conducted a survey of edge-on protoplanetary disks and found that the typical scale height of the millimeter sized dust was on the order of a few au at a radial distance of 100 au, compared to the gas in those systems which has a typical scale height of 10 au at a radius of 100 au. Thus, the scale height of the dust and gas in the primordial dust and gas scenario are expected to be different. Here, however, we find that the scale height of the gas and millimeter-sized dust is consistent with each other, which is not what is found in protoplanetary disks with primordial gas. Our finding that the millimeter-sized dust scale height is potentially correlated with the estimated \ce{CO} scale height for $\beta$ Pic, HD 32297, HD 110058, and HD 131488 could be consistent with the scenario that the dust and the \ce{CO} are of similar origin and produced in the same location in the disk, although it does not prove secondary origin of the gas. 

We also looked to see if the scale height of the millimeter dust in these four systems differs from the scale height we derive for \ce{CO}. We found that the scale height of the dust and \ce{CO} are consistent with each other due to the 20-60$\%$ error bars for the estimated \ce{CO} scale height. Having a better constraint on the inclination and radial distribution of the gas in these disks would give a more precise estimation of the \ce{CO} scale height. For now, we cannot conclude that the \ce{CO} and millimeter-sized dust have a different vertical distribution in these four systems.

\subsection{Is scale height an indicator of secondary origin gas?}
The 49 Ceti debris disk is also known to have \ce{CO} gas from ALMA \citep{moor19} and absorption from atomic carbon, but no absorption is seen from \ce{CO} \citep{roberge14}.  One  interpretation of the non-detection of CO in absorption towards 49 Ceti is that the scale height of the gas disk is not large enough to overcome the disk inclination. \cite{hughes18} suggests that the small scale height of the 49 Ceti gas is because the mean molecular weight is larger than what is expected from a primordial gas disk dominated by \ce{H2}.  

The HD 32297 debris disk differs from 49 Ceti in that it is more edge-on with an inclination of $\sim$88 degrees compared to 79 degrees \citep{hughes17}. Thus, the non-detection of \ce{CO} in absorption for HD 32297 is more constraining on the upper limit of the CO scale height. The small scale height upper-limit we find for HD 32297 could be due to a high mean molecular weight gas disk, which would support the secondary origin hypothesis as suggested in \cite{hughes18}. However, a recent study showed that the gas scale height might not be a useful probe of the mean molecular weight of the gas in a disk \citep{marino22}. 

Theoretical modeling done by \cite{marino22} found that if vertical mixing in the disk is weak, the distribution of \ce{CO} can be biased towards the midplane, due to the higher shielding from UV photons at the midplane of the disk where \ce{CO} will have a longer lifetime. The gas being distributed near the midplane of the disk will result in a small \ce{CO} scale height that may not be representative of the true distribution of gas in the system. Thus, the similarity we find between the estimated CO scale height and the millimeter dust scale height might not be due to a similar origin as the dust.

\subsection{Scale height in scattered light images}

The models presented in \cite{olofsson22} find that the presence of second generation gas in a debris disk of mass 0.1-1$M_{\oplus}$ can cause the micron-sized dust particles to settle towards the midplane as they migrate outwards from the planetesimal birth ring, while the millimeter-sized particles remain vertically extended near the birth ring. This is because the collisional timescales between dust grains is larger at further distances from the star. Therefore, at larger distances, dust grains are collisionally destroyed at a slower rate. Under the influence of gas drag, the micron-sized grains migrate outwards, where they can survive longer and will eventually settle towards the midplane as gas drag dampens their inclinations. The millimeter-sized grains remain near the planetesimal birth ring, where they are collisionally destroyed before their inclinations can be significantly dampened by gas drag. These models predict that debris disks with a CO mass in the range $\sim$0.1-1$M_{\oplus}$ should have larger millimeter-sized dust scale height (probed with ALMA images) than micron-sized dust scale heights (probed with scattered light images). 

We find that GPI does not vertically resolve the HD 131488 debris disk in H-band. Similarly, \cite{Kasper15} does not vertically resolve the HD 110058 debris disk in scattered light with SPHERE in H-band and reports a scale height upper limit of $<$4 au. We place an upper limit of $<$6 au for the scale height of HD 131488 in scattered light. For these two systems, the millimeter dust scale height is 9$^{+5}_{-4}$ for HD 131488 and 13$^{+5}_{-4}$ for HD 110058. Because of the degeneracy with scale height and inclination from the forward modeling, the ALMA dust scale height for HD 131488 is consistent with an upper limit if the inclination is less than 84 degrees. For HD 110058, and HD 131488 if the inclination is greater than 84 degrees, the millimeter-sized dust scale height is larger than the micron-sized dust scale height. 

 HD 110058 and potentially HD 131488  are well described by these models presented in \cite{olofsson22} because they have CO masses in the range of $\sim$0.1-1$M_{\oplus}$ and may have vertically extended millimeter-sized dust and vertically compact micron-sized dust. Additionally, the models of \cite{olofsson22} also show that if the gas is of primary origin, dominated by \ce{H2} and with a gas mass greater than 1$M_{\oplus}$, the disk will appear flat in both scattered light and millimeter images and there will not be a significant difference in the scale heights of the different dust sizes. HD 110058 and HD 131488 being potentially vertically resolved in ALMA continuum images and not vertically resolved in scattered light images is consistent with the secondary origin gas scenario presented by \cite{olofsson22}. 

\section{Conclusions}
We present high-spectral resolution (R=$\sim$60,000) M-band iSHELL observations to search for circumstellar CO absorption towards HD 32297 as well as GPI H-band polarimetry mode observations of the HD 131488 debris disk in scattered light. Our main findings are:
\begin{itemize}
    \item We do not detect any CO absorption with iSHELL and we measure a 3-$\sigma$ column density upper limit of $<6\times10^{15}$ cm$^{-2}$. We 
 estimate the scale height of the CO disk to be $\lesssim$ 2 au across its radial extent.
    \item We find a potential correlation between the millimeter dust scale heights measured from ALMA and the estimated gas scale heights in four systems: HD 32297, $\beta$ Pic, HD 110058, and HD 131488. This finding is consistent with the millimeter-sized dust and the CO having the same secondary origin from collisions with planetesimals.
    \item The estimated gas scale height for $\beta$ Pic is too large to be explained by the measured excitation temperature of the gas alone. This large scale height could be created by dynamical excitation from planetary mass objects that excite minor bodies 
    into inclined orbits.
    \item We find that HD 131488 is not vertically resolved in scattered light with GPI at H-band. The vertical structure of this disk seen in scattered light and potentially thermal emission is consistent with the models of \cite{olofsson22}, where the micron-sized grains settle to the disk midplane due to interactions with gas while the millimeter-sized grains remain on inclined orbits.
\end{itemize}

KW and CC acknowledge the support from the STScI Director's Research Fund  (DRF) and the NASA FINESST program. This work is supported by the National
Aeronautics and Space Administration under Grant No.
80NSSC22K1752 issued through the Mission Directorate. CL acknowledges support from the National Aeronautics and Space Administration under Grant No. 80NSSC21K1844 issued through the Mission Directorate. I.R. is supported by grant FJC2021-047860-I financed by MCIN/AEI /10.13039/501100011033 and the European Union NextGenerationEU/PRTR. AB and LM acknowledge research support by the Irish Research Council under grants GOIPG/2022/1895 and IRCLA/2022/3788, respectively. This paper makes use of the following ALMA data: ADS/
JAO. ALMA$\#$2018.1.00500.S, ADS/JAO.ALMA$\#$2015.1.01243. ALMA is a partnership of ESO
(representing its member states), NSF (USA) and NINS
(Japan), together with NRC (Canada) and NSC and ASIAA
(Taiwan), in cooperation with the Republic of Chile. The Joint
ALMA Observatory is operated by ESO, AUI/NRAO and
NAOJ. The National Radio Astronomy Observatory is a
facility of the National Science Foundation operated under
cooperative agreement by Associated Universities, Inc.

\software{
This research has made use of the following software projects:
    \href{https://astropy.org/}{Astropy} \citep{astropy18},
    \href{https://matplotlib.org/}{Matplotlib} \citep{matplotlib07},
    \href{http://www.numpy.org/}{NumPy} and \href{https://scipy.org/}{SciPy} \citep{numpy07},
    CASA\citep{CASA}
    and
    the NASA's Astrophysics Data System.
}

\section*{\textbf{Appendix A}}
To estimate the scale height of the gas in these four disks, the inclination, CO mass, and CO radial distribution must be known. For HD 110058, the outer radius is not reported in the literature. In order to perform the analysis of the scale height of HD 110058 we first use publically available data cubes from the ALMA archive of $^{12}$CO and  \citep{Hales2022} to measure the outer radius of the gas. 

\begin{figure}[!htbp]
    \centering
    \includegraphics[scale=0.6]{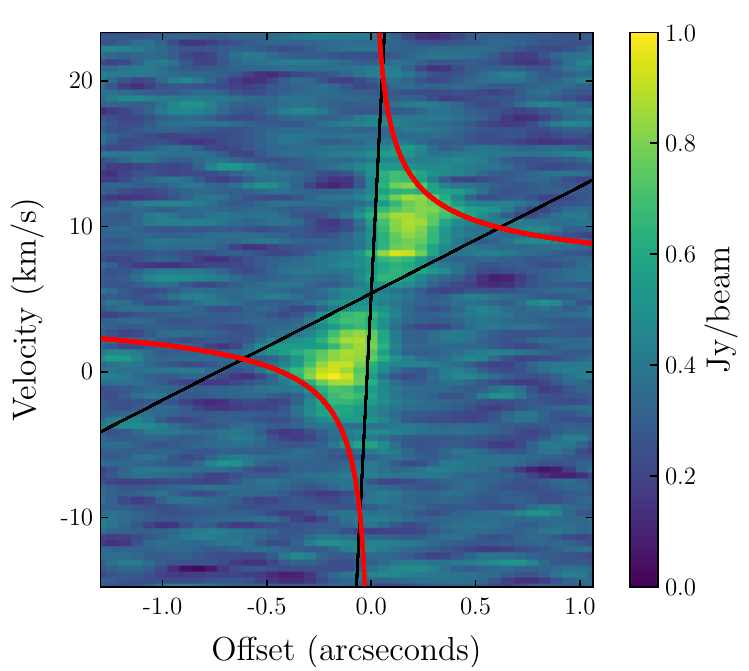}
    \caption{PV diagram of $^{12}$CO J=3-2 emission around HD 110058. The red lines show the Keplerian orbit with an inclination of 85.5 degrees and a stellar mass of 1.84 $M_{\odot}$. The black lines show different radii in the orbital plane of the disk, corresponding to an R$_{in}$ of 7 AU and an R$_{out}$ of 80 AU. }
    \label{fig:PVdiagram}
\end{figure}

To determine the outer radius of the gas disk around HD 110058, we use the $^{12}$CO data cube from the ALMA archive \citep{Hales2022} to create a position-velocity (PV) diagram. We construct a PV diagram across the horizontal axis of the disk. Diagonal lines on the PV diagram represent different orbital radii in the plane of the disk and we fit diagonal lines to the outer edge of the PV diagram to determine the disk radii. We use a stellar mass of 1.84 M$_{\odot}$ and an inclination of 85.5 degrees. The inclination and stellar mass is derived from simultaneously fitting the $^{12}$CO J=3-2 and J=2-1 and $^{12}$CO J=3-2 lines with a radiative transfer model in \cite{Hales2022}. To fit the PV diagram, we first determine its outer edge. This is done by finding, at every offset position, the velocity where the flux drops below a 5$\sigma$ threshold, like was done in \cite{seifried16}. If there is no flux value above the 5$\sigma$ threshold for a given offset position, we do not include that offset position in the fitting. The PV diagram with the best fit radii (black lines) is shown in Figure \ref{fig:PVdiagram}. The two black diagonal lines in Figure \ref{fig:PVdiagram} represent the best fit orbital radii of 7 and 80 AU to the edges of the PV diagram. The red curve shows a Keplerian orbit with the stellar mass and disk inclination reported by \cite{Hales2022}. With this fit we constrain the radial distribution of the gas to be between 7$\pm$7 and  80$\pm$10 AU. This measurement for the outer radius is used in the calculation of the scale height of HD 110058. 

\bibliographystyle{aasjournal}
\bibliography{mybib}






\end{document}